\documentclass[bibyear]{aa} % if the references are not structured ...quello che sto usando
\usepackage{graphicx}
\usepackage{amsmath}
\usepackage{txfonts}
\def\ltsim{\; \raise0.3ex\hbox{$<$\kern-0.6em \raise-1.1ex\hbox{$\sim$}}\; }
\def\gtsim{\; \raise0.3ex\hbox{$>$\kern-0.75em \raise-1.1ex\hbox{$\sim$}}\; }

\def\ie{{\it i.e.,~}}

\def\eg{{\it e.g.,~}}

\begin{document} 
\title{Can giant radio halos probe the merging rate of galaxy clusters?}
%   \subtitle{I. Overviewing the $\kappa$-mechanism}
   \author{R. Cassano
          \inst{1}
          \and G. Brunetti
          \inst{1}
          \and C. Giocoli
          \inst{2}
          \and S. Ettori
          \inst{3,4}
          %\and
          %C. Ptolemy\inst{2}\fnmsep\thanks{Just to show the usage
          %of the elements in the author field}
          }
 \institute{INAF-Istituto di Radioastronomia,
              via P. Gobetti, 101, 40129, Bologna, Italy\\
              \email{rcassano@ira.inaf.it}
         \and
         	Aix Marseille Universit\'e, CNRS, LAM (Laboratoire d'Astrophysique de Marseille) UMR 7326, 13388, Marseille, France
	 \and
             INAF/Osservatorio Astronomico di Bologna, via Ranzani 1, I--40127 Bologna, Italy 
             \and
             INFN, Sezione di Bologna, viale Berti Pichat 6/2, I-40127 Bologna, Italy
                 }
%\date{Received September 15, 1996; accepted March 16, 1997}
%\date{Received 01/03/2016; accepted 20/06/2016}
% \abstract{}{}{}{}{} 
% 5 {} token are mandatory
 %
\abstract{
Observations of galaxy clusters both in the radio and X-ray bands probe a direct link between cluster mergers and giant radio halos, suggesting that these sources can be used as probes of the cluster merging rate with cosmic time. 
However, while all giant radio halos are found in merging clusters not every merging cluster host a giant radio halo. In this paper we carry out an {\it explorative} study that combines the observed fractions of merging clusters and radio halos with the merging rate predicted by cosmological simulations and attempt to infer constraints on merger properties of clusters that appear disturbed in X-rays and of clusters that host radio halos. 
We use classical morphological parameters to identify merging systems and analyze the currently largest (mass-selected $M_{500}\gtrsim 6\times10^{14} \,M_\odot$ and $0.2\leq z\leq0.33$) sample of galaxy clusters with radio and X-ray data; we extract this sample from the Planck Sunyaev-Zeldovich cluster catalogue. We found that the fraction of merging clusters in this sample is $f_m\sim 62-67\%$ while that of clusters with radio halos is $f_{RH}\sim 44-51\%$. 
We assume that the morphological disturbance measured in the X-rays is driven by the merger with the largest mass ratio, $\xi$ ($\xi=M_i/M_1<1$ with $M_i$ and $M_1$ being the progenitor masses), that is still ongoing in the cluster at the epoch of observation. 
Results from theoretical studies allow to derive the fraction of mergers with mass ratio above a minimum threshold (those with $\xi\gtsim\xi_{min}$) in our sample, under the assumption of a timescale $\tau_m$ for the duration of merger-induced disturbance. The comparison of the theoretical merger fraction with the observed one allows to constrain a region in the ($\xi_{min}$,$\tau_m$) plane. 
We find that under the assumption of $\tau_m\sim 2-3$ Gyr, as constrained by simulations, the observed merger fraction matches the theoretical one for $\xi_{min}\sim 0.1-0.18$. This is consistent with optical and near-IR observations of galaxy clusters in the sample that constrain $\xi_{min}\simeq 0.14-0.16$ through weak lensing analysis or study of the velocity distribution of galaxies in the clusters. 
The fact that radio halos are found only in a fraction of merging galaxy clusters may suggest that merger events generating radio halos are characterized by larger mass ratio; this seems supported by optical/near-IR observations of RH clusters in the sample that indeed allow to constrain $\xi_{min}\sim0.2-0.25$. Alternatively, radio halos may be generated in all mergers but their lifetime is shorter (by $\sim f_{RH}/f_m$) than the timescale of the merger-induced disturbance.
We stress that this is an {\it explorative} study, however it suggests that follow up studies using the forthcoming radio surveys and adequate numerical simulations have the potential to derive quantitative constraints on the link between cluster merging rate and radio halos at different cosmic epochs and for different cluster masses.
}
\maketitle
%
%________________________________________________________________
%

\section{Introduction}

%Galaxy clusters are of great interest for modern cosmology since they are the largest and more recently assembled structures in the Universe. 
In the paradigm of the hierarchical structure formation scenario, galaxy clusters, the largest and more recently assembled structures
in the Universe, form via mergers of smaller halos and continuous accretion of unbound matter. 
The process of mass accretion of dark-matter halos is a clear outcome of the cosmological model. It can be statistically investigated
with N-body simulations and semi-analytical models through the identification of merger trees of dark matter halos, which lead to the derivation of the mass accretion history 
and merging rate as a function of redshift, halo mass and mass ratio of the progenitors 
%mass accretion rate (MAR) as a function of redshift, mass ratio of the sub-halos and mass of the progenitors 
(\eg van den Bosh 2002; Giocoli et al. 2007; Moreno et al. 2008; Fakhouri \& Ma 2008; 
McBride et al. 2009; Fakhouri et al. 2010; Giocoli et al. 2012). 

%(van den Bosh 2002; Zhao et al. 2003; Sheth \& Tormen 2004a,b; Giocoli et al. 2007; McBride et al. 2009; Zhao et al. 2009).
%In particular, this is investigated with the identification of merger trees of dark matter halos, which lead to the derivation of the mass accretion history 
%and mass accretion rate (MAR) as a function of redshift, mass ratio of the sub-halos and mass of the progenitors (\eg van den Bosh 2002; Fakhouri \& Ma 2008;  
%McBride et al. 2009; Fakhouri et al. 2010; Giocoli et al. 2012). 

Observationally, the exploration of the merging rate  
%MAR of individual 
of dark matter haloes has only been attempted on the scales of galaxies by using two main methods for tracing the merging history in the observations: morphological identification techniques (Conselice et al. 2003;
Lotz, Primack \& Madau 2004) and the close galaxy pair method (\eg Patton et al. 2000; De Propris et al. 2005).
These methods are then combined with the merger time-scale derived from N-body simulations to get the merging rate (\eg Lotz et al. 2011; Jian et al. 2012; Conselice 2014).
%combining the number of observed
%pairs of close or disturbed galaxies with the merger probability and time-scale derived from N-body simulations (\eg Lotz et al. 2011; Jian et al. 2012; Conselice 2014). 
However, current results are inconclusive, because the merger rate of dark-matter halos and the merger rate of galaxies do not necessary coincide because 
they are related by dissipative processes (dynamical friction, tidal interaction, stellar feedback, etc.) that are difficult to model 
(Fakhouri \& Ma 2008; Guo \& White 2008; Lotz et al. 2011; Hopkins et al. 2013). Dissipative processes are instead less relevant during the mass accretion of galaxy clusters. 
Nevertheless only recently a method 
%to derive the mass accretion rate of galaxy clusters has been proposed and it is 
based on the possibility to measure 
 the cluster mass in a thin spherical shell surrounding the cluster beyond $R_{200}$ (with the caustic technique) and by estimating its infalling time (\eg Diaferio 2015; De Boni et al. 2016), have been proposed to measure the mass accretion rate of galaxy clusters.
%no measure of the MAR of galaxy cluster have been attempted so far. This is due to the observational difficulties of estimating the cluster merger rate. Very recently, a novel method based on the measuring of the mass of spherical shell surrounding the cluster (with the caustic technique) and by estimating its infall time (De Boni et al. 2015; Diaferio 2015) has been proposed. 
However, in general, the growth of structures on the scale of galaxy clusters remains poorly explored from an observational prospective (\eg Lemze et al. 2013).

Mergers between clusters are the most energetic phenomena since the big-bang, with a release of a gravitational potential energy of $\sim10^{63}-10^{64}$ ergs 
during one cluster crossing time ($\sim 1$ Gyr). During such events shock waves and random vortical flows, if not turbulence, are produced in the intracluster medium (ICM) (\eg Kulsrud et al. 1997; Norman \& Bryan 1999; Ricker \& Sarazin 2001). These motions originate due to vorticity generation in oblique accretion shocks and instabilities during the cluster formation, and in the wakes of the smaller subclusters (\eg Subramanian et al. 2006; Brunetti \& Jones 2014; Br\"uggen \&Vazza 2015). The bulk of the gravitational energy associated with the collision will be released as thermal energy in the final system (\eg Kravtsov \& Borgani 2012), while another fraction can be channeled into non-thermal plasma components, \ie relativistic particles and magnetic fields in the ICM (\eg Brunetti \& Jones 2014). 
The existence of cosmic ray electrons and magnetic fields in the ICM is in fact demonstrated by radio observations. Cluster-scale ($\sim$Mpc-scale) diffuse synchrotron emission is frequently found in merging galaxy clusters in the form of so-called giant radio halos (hereafter RH), apparently unpolarized synchrotron emission associated with the cluster X-ray emitting regions, and giant radio relics, elongated and often highly polarized synchrotron sources typically seen in the clusters outskirts (\eg Feretti et al. 2012, for an observational review). The properties of radio relics suggest a connection with large scale shocks that cross the ICM during mergers and that may accelerate locally injected electrons or reaccelerate pre-existing energetic electrons, while RH likely trace gigantic turbulent regions in the ICM, where relativistic electrons can be reaccelerated through scattering with MHD turbulence (\eg Brunetti \& Lazarian 2007; Br\"uggen et al. 2012).

In the last decades, radio observations of statistical samples of galaxy clusters have shown that RHs are not ubiquitous, only $\sim 20-30$\% of the X-ray luminous ($L_X(0.1-2.4\,\mathrm{keV})\geq5\times10^{44}$ erg/s) clusters host a RH (\eg Venturi et al. 2008; Kale et al. 2015), while the fraction of clusters with RH becomes larger in SZ-selected clusters (\eg Basu 2012; Cassano et al. 2013; Sommer \& Basu 2014; Cuciti et al. 2015). Most important, it was found that RH and non-RH clusters are clearly separated in the $P_{1.4}-L_X$ and $P_{1.4}-M_{500}$ ($Y_{500}$) diagrams according to the cluster dynamical status, with RH always associated to dynamically disturbed clusters and clusters without RHs statistically more ``relaxed'' (\eg Brunetti et al. 2007, 2009; Cassano et al. 2010, 2013). The connection between RHs and merging clusters has been further supported by a number of independent studies (\eg Rossetti et al. 2011; Wen \& Han 2015; Parekh et al. 2015; Mantz et al. 2015; Yuan et al. 2015; Kale \& Parekh 2016). 
The RH--cluster merger connection suggests that RHs can be used as signposts of cluster mergers and support the idea that RH are ``transient'' phenomena tracing turbulent region in the ICM during the process of cluster formation. However not all merging clusters host a giant RH (see Cassano et al. 2013 and ref. therein) and this poses fundamental questions about the conditions that are necessary to generate cluster-scale synchrotron diffuse emission.

Interestingly the connection between mergers and non-thermal phenomena also opens to the possibility to infer constraints on the cluster merging rate from radio observations. 
In this paper we start exploring this possibility. In particular we attempt to combine the observed fraction of merging clusters and the observed fraction of RH in clusters with the merging rate predicted by cosmological simulations to infer constraints on the properties of the mergers that induce disturbances observed in the X-rays and of those responsible for RH.

We stress that this is an explorative study, with the main aim to start to investigate the possibility to use diffuse radio emission in galaxy clusters as tracer of the cluster dynamical status. 
%To do this we use the currently most complete mass-selected sample of galaxy clusters with radio and X-ray information. 
In particular, we stress that current statistical information is still limited to very massive ($M_{500}\gtsim 6\times 10^{14}\, M_{\odot}$) and relatively nearby systems ($z\simeq 0.2-0.33$), while we anticipate that better constraints can be obtained using less massive systems or clusters at higher redshifts.

In Sect.2 we present the cluster sample and derive the fractions of merging clusters and that of clusters with RH; in Sect.3 we describe the formalism by Fakhouri, Ma \& Boylan-Kolchin (2010) to derive the merging rate in simulations and derive the expected merger fraction. In Sect.4 we compare the observed merger fraction and RH fraction with expectations from simulations and attempt to constrain the properties that cluster mergers should have to explain the observed fraction of clusters with X-ray disturbances and that of clusters with RH. Finally, in Sect.5 we summarize the main results and discuss the main implications for the origin of giant RH in galaxy clusters.

A $\Lambda$CDM cosmology ($H_{o}=70\,\rm km\,\rm s^{-1}\,\rm Mpc^{-1}$, $\Omega_{m}=0.3$, $\Omega_{\Lambda}=0.7$) is adopted.

%__________________________________________________________________

\section{Data and sample selection}
\label{sample}

We used the Planck Sunyaev-Zeldovich (SZ) cluster catalogue (PSZ, Planck Collaboration XXIX 2014a) to select 54 clusters with $M_{500}\gtrsim 6\times10^{14} \,M_\odot$\footnote{The values of $M_{500}$ in the PSZ catalogue are obtained from $Y_{500}$ as described in Sect.7.2.2 in Planck Collaboration XXIX 2014a.}, redshift  $0.2\leq z\leq0.33$ and $\delta>-30^{\circ}$ and $|b|\geq \pm20^{\circ}$, where b is the galactic latitude (Tab.~\ref{Tab.A1}).
With such a selection the sample has a mass-completeness of $\sim80\%$\footnote{This completeness is estimated in $Y_{SZ}$ by Planck Collaboration (2014a) and then converted in "mass completeness" using scaling relations in Planck Collaboration XX (2014b; see Fig.28 in Planck Collaboration 2014a). }.
%, high enough for our  purpose.

This selection has been thought to optimize the available information in the radio band, indeed 37 out of 54 clusters belong to the Giant Metrewave Radio Telescope ({\it GMRT}) RH Survey and its extension (EGRHS; Venturi et al. 2007, 2008; Kale et al. 2013, 2015) and for 39 out of 54 clusters ($\sim 72\%$ of the sample) information about the presence/absence of diffuse radio emission is available. In particular, 17 clusters host giant RH, while 3 clusters host candidate RHs (see Tab.~\ref{Tab.A1}). The fraction of RH, defined as $f_{RH}=N_H/N_{tot}$, with $N_H$ being the number of RH and $N_{tot}$ the total number of clusters, is thus $\sim 44\%$ and can reach $\sim 51\%$ if we include the 3 uncertain cases. 

51 out of 54 clusters ($\sim 94\%$ of the sample, including all the 39 clusters with available radio information) have X-ray data (Chandra and/or XMM-Newton) that can be used to derive information about the cluster dynamical status. In particular, 41 of these clusters have Chandra data for which morphological indicators, such as the power ratio $P_3/P_0$ (\eg Buote \& Tsai 1995), the emission centroid shift $w$ (\eg Mohr et al. 1993) and the surface brightness concentration parameter $c$ (\eg Santos et al. 2008) can be homogeneously derived to quantitatively establish the cluster dynamical status.
Following Cassano et al. (2010, 2013), we adopted an algorithm for an automatic detection of the point sources, which are then removed from the images. We study the cluster substructures on a typical RH scale analyzing the surface brightness inside an aperture radius of 500 kpc, since we are interested in the cluster dynamics on the scales where the energy is most likely dissipated to generate the radio emission. We briefly remind that the power ratio is a multipole decomposition of the two-dimensional projected mass distribution within a given aperture, and $P_3/P_0$ is the lowest power ratio moment providing a clear substructure measure (B\"ohringer et al. 2010). The centroid shift $w$ is defined as the standard deviation of the projected separation between the peak and the centroid of the cluster X-ray brightness distribution in unit of the aperture radius. The concentration parameter $c$ is defined as the ratio of the peak (within $100$ kpc) over the ambient (within $500$ kpc) X-ray surface brightness. Following previous papers (Cassano et al. 2010, 2013; Cuciti et al. 2015), we adopt the following threshold values to classify clusters as mergers: $P_3/P_0\gtsim 1.2\times10^{-7}$, $w\gtsim 0.012$ and $c\ltsim0.2$. 
For 32 clusters with Chandra data morphological parameters are already published in Cassano et al. (2010, 2013) and Cuciti et al. (2015), here we derive the morphological quantities for additional 9 clusters following the approach outlined above (and described in previous works; see \eg Sect. 3 of Cassano et al. 2010 for details). The resulting dynamical status of the clusters, ``merger'' vs ``relaxed'', is reported in Tab.~\ref{Tab.A1} (column 6); the values of the morphological parameters for the 41 clusters with Chandra data are reported in Tab.\ref{Tab.B1}.

10 more clusters with available XMM-Newton observations can be added to this sample and after a visual inspection of their images we can assess (even if with less confidence) their dynamical status (also reported in Tab.~\ref{Tab.A1}, column 6). 

In deriving the merger fraction\footnote{To compare the observed merger fraction with theoretical expectations we convert the $M_{500}$ to virial masses, $M_{vir}$, by assuming a NFW profile (\eg Navarro, Frenk \& White 1997) for the dark matter halos and the concentration-mass relation in Duffy et al. (2008), see Appendix A in Ettori et al. (2010). Both the values of $M_{500}$ and $M_{vir}$ are reported in Tab.~\ref{Tab.A1}.}
 we assume that the disturbance we measure in the X-rays is mainly due to the merger with the largest mass ratio that is ongoing in the system at the epoch of the observation, \ie a binary merger approximation. Under this assumption the merger fraction is equivalent to the fraction of clusters that is actually in phase of merger (where with clusters we refer to the final product of the merger).
Considering the sample of 39 clusters with available radio information, we found that the fraction of dynamically disturbed systems, or the merger fraction, defined as $f_{m}=N_m/N_{tot}$ with $N_m$ being the number of merging clusters, is $\sim 62-67\%$ (including the uncertainty on the classification of two clusters; see Tab.~\ref{Tab.A1}).
If we extend this analysis to the sample of 51 clusters with X-ray data, we found $f_m\sim 65-69\%$. We can only speculate on the fraction of RH in this latter sample, for instance, by assuming that the fraction of merging clusters with RH in these additional 12 clusters is the same that we measure in the sample of 39 clusters (that is $\sim 70\%$), we obtain $f_{H}\sim 45-51\%$. The derived fractions are summarized in Tab.\ref{tab1}.
%and it becomes $\sim$ 65\% when also clusters with XMM-Newton data are included. 

\begin{table}[h]
\caption[]{Cluster's fractions}\label{tab1}
\begin{center}
\footnotesize
%\vspace*{-0.8 cm}
\begin{tabular}{ccc}
\noalign{\smallskip}
\hline\noalign{\smallskip}
%\hline\noalign{\smallskip}
cluster sample & $f_m$ & $f_{RH}$  \\
%			   &    & [$10^{14} M_{\odot}$] & [$10^{14} M_{\odot}$] & & & \\	
%\begin{table}[htbp]
%\begin{center}
%\begin{tabular}{l c c c c c c}
%\hline
%\hline
%cluster name &    RA    &      Dec&         z& $M_{500} ( 10^{14}\,M_\odot$)&  radio info& Xinfo\\
\hline
39 clusters & 62-67\%   & 44-51\% \\
51 clusters & 65-69\% & 45-51\% \\
\hline
%\end{tabular}
%\caption[high redshift sample cluster properties]{high redshift sample clusters porperties} \label{tab:highzsample}
%\end{center}
%\end{table} 
\noalign{\smallskip}
\hline\noalign{\smallskip}
\end{tabular}
\end{center}
\end{table}

Calculations in the paper will be based on the fractions extracted from the sample of 39 clusters (for which both radio and X-ray data\footnote{Note that for 4 out of 39 clusters the dynamical classification is based on a visual inspection of the XMM-Newton cluster image.} are available), although these fractions are not expected to change significantly in the extended sample (under reliable assumptions, see Tab.~1).

\section{Merging rate of halos from simulations}

In the Lambda cold dark matter ($\Lambda$CDM) scenario dark matter halos grow in mass and size primarily through mergers with other halos: merger with comparable mass halos (``major mergers''), merger with smaller satellite halos (``minor mergers'').
% and accretion of non-halo material. 
To derive the merging rate of halos we use the result derived from the combined Millennium (Springel et al. 2005) and Millennium-II (Boylan-Kolchin et al. 2009) simulations (Fakhouri, Ma \& Boylan-Kolchin 2010; FMB10 hereafter). FMB10 used merger trees of dark matter halos to extract a catalog of mergers containing for each descendent halo at redshift $z_d\geq0$ with mass $M$ the $N_p$ ($N_p\geq1$) progenitors at $z_p=z_d+\Delta z$, with masses $M_1\geq M_2\geq...M_{N_p}$. To derive the merging rate they include {\it all} the progenitors (above a given mass threshold), and since they do not have information about the order the progenitors merge, they assume that each progenitor $M_i$ with $i\geq 2$ mergers with the most massive progenitor $M_{1}$ at a given point between $z_p$ and $z_d$. Thus a descendent halo with $N_p$ progenitors is assumed to be the result of $N_p-1$ binary merger events with mass ratio $\xi= M_i/M_1\leq 1$ ($i=2,...N_p$). The progenitor mass ratio, $\xi$, is defined so that, for instance, $\xi=0.3$ indicates major mergers (with mass ratio $1:3$), while $\xi=0.1$ indicates mergers with mass ratio $1:10$. The derived mean merging rate per halo, $dN_m/d\xi/dz$, that gives the mean number of mergers per unit halo per unit $z$ per unit $\xi$, can be well described by the following formula (FMB10):

% with $\xi$ being the progenitor mass ratio. $\xi=M_i/M_1\leq 1$ (with $i=1, N_p$) is defined so that, for instance, $\xi=0.3$ indicates major mergers (with mass ratio $1:3$), while $\xi=0.1$ indicates mergers with mass ratio $1:10$.  FMB10 found that the mean merging rate per halo can be well fitted by the following equation:

 \begin{equation}
 \label{Eq.MR}
\frac{dN_m}{d\xi dz} (M,\xi,z)=A\,\Bigg(\frac{M}{10^{12}M_{\odot}}\Bigg)^{\alpha}\,\xi^{\beta}\exp\Bigg[\Bigg(\frac{\xi}{\tilde{\xi}}\Bigg)^\gamma \Bigg]\,(1+z)^\eta
\,,
   \end{equation}

\noindent were the best-fit parameters are: $\alpha=0.133$, $\beta=-1.995$, $\gamma=0.263$, $\eta=0.0993$ and $A=0.0104$, $\tilde{\xi}=9.72\times10^{-3}$. This formula has a negligible dependence on the redshift and it is also nearly independent of the mass. The main dependence is on the mass ratio $\xi$, so that the number of merger per halo is larger for smaller mass ratio. For example, the number of mergers (per halo) with $\xi=0.01$ is about 90 times larger than the number of mergers with $\xi=0.1$ (see Fig.1, right panel, in FMB10). 

Integrating Eq.~\ref{Eq.MR} for $\xi\geq \xi_{min}$ and between $z_0$ and $z$ one obtains the cumulative number of mergers, $N_m(\xi_{min}, M_0, z_0, z)$, that is the total number of mergers with $\xi\geq \xi_{min}$ that a halo of mass $M_0$ at redshift $z_0$ has encountered between $z_0$ and an earlier $z$ during the halo's history:

 \begin{equation}
 \label{Eq.NM}
N_m(\xi_{min}, M_0, z_0, z)=\int_{z_0}^{z} dz\, \int_{\xi_{min}}^{1}d\xi\frac{dN_m}{d\xi dz}(M(z),\xi,z)
\,,
   \end{equation}

\noindent where $M(z)$ is the mass accretion history and can be obtained integrating the fitting formula for the mass accretion rate, $\dot{M}=(M_0-M_1)/\Delta_t$ (where $M_0$ is the descendent mass at time $t$ and $M_1$ is the mass of its most massive progenitor at time $t-\Delta_t$), that is given by (FMB10):

\begin{equation} \label{Eq.MAR}
%\begin{split}
\langle\dot{M}\rangle_{mean}  =46.1\,\mathrm{M_{\odot} yr^{-1}} \Bigg(\frac{M}{10^{12}\,M_{\odot}}\Bigg)^{1.1}\times E_z\,(1+1.11\,z) 
\,,
\end{equation}

%\noindent Since Eq.\ref{Eq.NM} provide the number of merger per halo, to derive the fraction of merging cluster to be compared with the observations we solve Eq.~\ref{Eq.NM} for each cluster of the sample with mass $M_0$ at $z_0$ as a function of different values of $\xi_{min}$ and 

\begin{figure}
   \centering
\includegraphics[width=8cm]{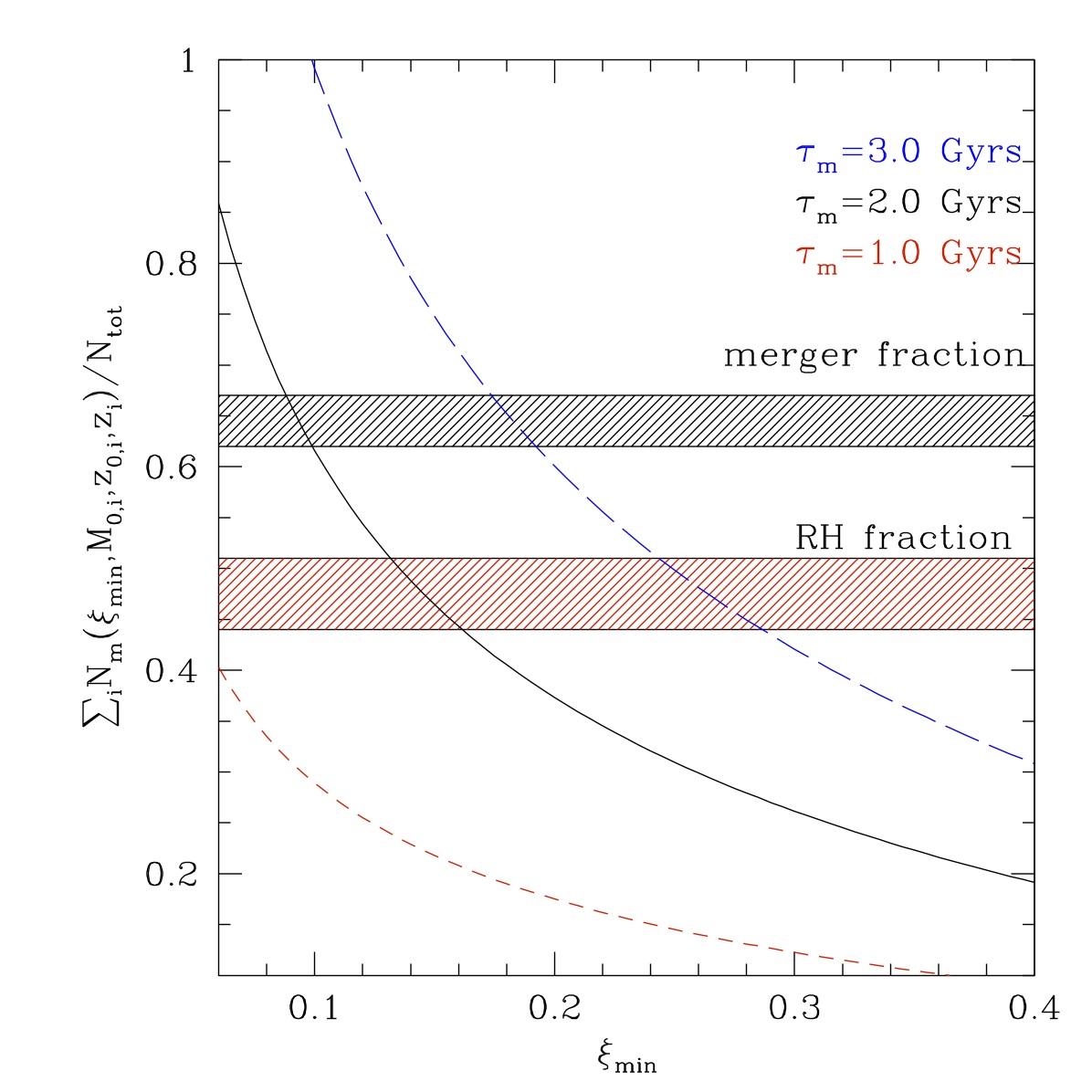}
%{mergerhalo_fraction_35_51.ps}
      \caption{Predicted average merger fraction  for clusters in the sample as a function of $\xi_{min}$, assuming three different values for the merger timescale $\tau_m=1, 2, 3$ Gyr (from bottom to top). The observed merger fraction and RH fraction are also reported (shadowed regions).}
         \label{Fig.mergerfraction}
   \end{figure}

The quantity we derived observationally is the merger fraction, that is the fraction of clusters with significant dynamical disturbance in the X-rays (see Sect.~2). 
To compare this quantity with expectations given by Eqs.~(1)-(3),
we need to assume a merger timescale, $\tau_m$, associated with the duration of the morphological disturbance that we infer from X-ray images. 
%Numerical simulations of cluster mergers can be used to put constrain on $\tau_m$ (\eg Tormen et al. 2004; Poole et al. 2006; Hallman \& Jeltema 2011), however we prefer to leave it as a free parameter and discuss in Sect.4 the implications of particular choice of $\tau_m$ on our results.
This is a free parameter in our calculations that however can be constrained through numerical simulations (see Sect.~4).
We derive the average fraction of mergers with $\xi\geq \xi_{min}$ expected in our sample by integrating  Eq.~\ref{Eq.NM} for each cluster of the sample with mass $M_0$ and redshift $z_0$ up to the redshift $z$ corresponding to the look back time $\tau_m$ and then computing $\sum_i N_m(\xi_{min}, M_{0,i}, z_{0,i},z_{i})/N_{tot}$, where the sum is on the $N_{tot}$ number of clusters in the sample.
%integrate Eq.~\ref{Eq.NM} in $dz$ up to the redshift $z$ corresponding to $\tau_m$, as a function of $\xi_{min}$. Then the average merger fraction is  $\sum_i N_m(\xi_{min}, M_{0,i}, z_{0,i},z_{i})/N_{tot}$, where the sum is on the $N_{tot}$ number of clusters in the sample.
The derived average merger fraction is reported in Fig.~\ref{Fig.mergerfraction} as a function of $\xi_{min}$ for three values of $\tau_m$ ( $\tau_m=1, 2, 3$ Gyr) and is compared with the observed merger fraction and the observed fraction of clusters with RH (shadowed regions).  

The predicted merger fraction decreases for larger mass ratios simply because major mergers are less common than minor ones, and it obviously increases by assuming larger timescales.

%------------------------figure 1----------------------------------------------------
%-------------------------------------------------------------------------------------

\section{Comparison with theory}

The comparison between the ``observed'' and ``theoretical''  merging fractions allows to derive constraints on relevant parameters, such as $\tau_m$ and $\xi_{min}$. 
Before proceeding in this direction, we need to discuss some caveats in our procedures. In principle, the comparison between the ``observed'' and ``theoretical''  merging fractions allows to derive constraints on relevant parameters, such as $\tau_m$ and $\xi_{min}$. 

\subsection{Caveats}

The observed merger fraction is derived by measuring the fraction of clusters with significant X-ray disturbances. This means that our method is limited to events with significant mass accretion, otherwise it would be difficult to classify these events as ``mergers'' based on the morphological parameters. The dynamical parameters are derived within a region of radius $500$ kpc.
%, as a consequence, observationally, the merger fraction is derived by considering as merging clusters those with morphological disturbances in their central $1$ Mpc (diameter) region. 
On the other hand, in FMB10 the merger fraction is derived from the merging rate which considers all the infalling halos within the virial radius of the main cluster (that for our clusters is $\sim 2-3$ Mpc). However, since we are considering a ``rate'', number of infalling halos per unit time, what is important is that the halos crossing the virial radius of the main cluster cross, at a given time, the radius of 500 kpc reaching the central regions. According to cosmological simulations, halos with a mass ratio $\xi\gtsim 0.1$ reach their pericentric distance, that is $\sim 0.2-0.3 R_v$, within a time-scale of  $\sim 0.9$ Gyr from the virial crossing (\eg Tormen et al. 2004).

The other assumption is that of binary mergers, \ie we assume that the disturbance we measure in the X-rays is mainly caused by a binary merger event between the two main progenitors. Consequently, we derive the expected average merger fraction from the fitting formulae by FMB10 assuming that in a merger timescale there is a main binary merger event with $\xi\gtsim \xi_{min}$ that influence the observed dynamical status . However, in a merger timescale, especially for long timescale, clusters might experience multiple merger episodes characterized by lower mass ratio ($\xi<\xi_{min}$). If the number of mergers with
a mass ratio slightly smaller than $\xi_{min}$ is significant, our assumption would be no longer valid, since also the interplay of these mergers would contribute to the
morphological disturbance. 
%. However, we show below that the contribution of these slightly minor mergers is expected to be marginal. 

%------------------------figure 2----------------------------------------------------
   \begin{figure}
   \centering
\includegraphics[width=8cm]{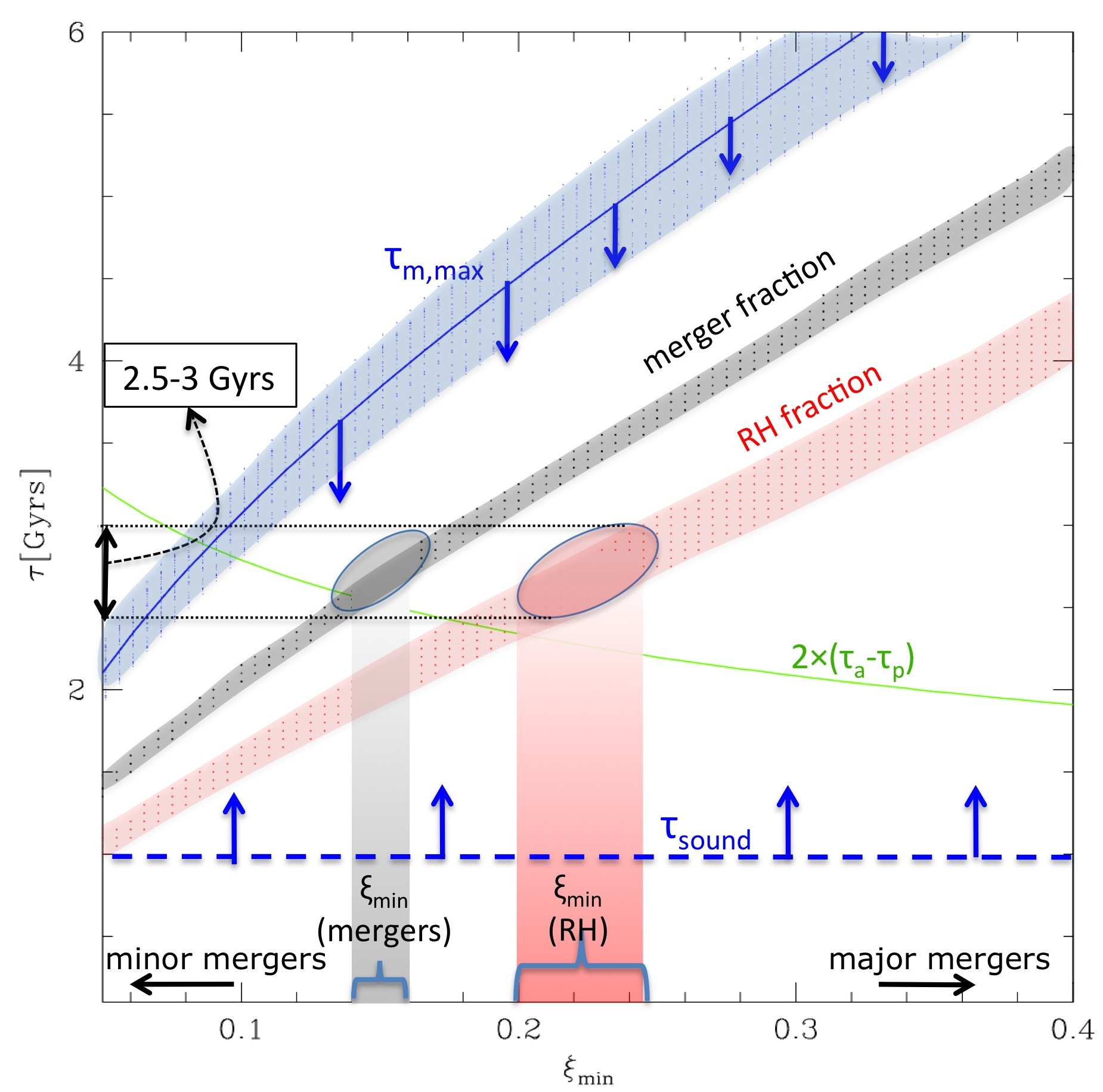}
%{mergerhalo_fraction_35_51.ps}
      \caption{Allowed regions of parameters ($\xi_{min}$, $\tau_m$) constrained by requiring that the observed merger fraction (black region) and RH fraction (red regions) match those predicted by theory (these regions account for the uncertainty in the observed fraction). For each $\xi_{min}$ the maximum allowed value of $\tau_m$, $\tau_{m,max}$ (blue region with arrows; see Sect.~4.2) and a lower limit to $\tau_m$ (horizontal blue dashed line, see Sect.~4.2) are also shown. The merger-timescale constrained by cosmological simulation is also reported (green line; Tormen et al. 2004). Ranges of $\xi_{min}$ constrained through optical/near-IR observations of galaxy clusters in the sample (see Sect.4.4) are shown for merging clusters (black rectangular) and for clusters with RH (red rectangular). These values of $\xi_{min}$ both constrain $\tau_{m}\sim\tau_{RH}\sim 2.5-3$ Gyr (this is also show in the Figure).}
         \label{Fig.tau_ximin}
   \end{figure}
%-------------------------------------------------------------------------------------

\subsection{Mergers and observed X-ray morphology}

Given these premises, we proceed with the comparison between the observed merger fraction and the results by FMB10. From Fig.~\ref{Fig.mergerfraction} it is clear that we can find combinations ($\tau_m$, $\xi_{min}$) 
for which the expected merger fraction can match that observed in our sample. Fig.\ref{Fig.tau_ximin} (black region) shows the allowed regions of ($\tau_m$, $\xi_{min}$) that is derived by matching theory and observations.
% derived allowed region in the ($\tau_m$, $\xi_{min}$) diagram;  
% These constraints are derived to match both the observed total merging rate per halo (black region) and the observed merging rate of clusters with radio halos (red region). 
As expected, there is a clear degeneracy between $\tau_m$ and $\xi_{min}$. 

We first identify forbidden regions in the $\tau_m$-$\xi_{min}$ diagram.
A lower limit on $\tau_m$ can be obtained by assuming that the merger-driven perturbations within a region of diameter $1$ Mpc cannot last for a timescale shorter than the sound crossing time of that region, that for a galaxy cluster with $T\sim10^8$ K is $\tau_{sound}\simeq 1$ Gyr. 
We note that values of $\tau_{m}$ constrained by the observations are always larger than $\tau_{sound}$ (at least for $\xi_{min}\geq 0.05$). 
An upper bound to $\tau_m$ can be derived by considering the fact that extremely large duration of mergers would make dynamically disturbed all the clusters that are observed at a given cosmic epoch. 
Specifically, for each cluster of the sample with mass $M_0$ at redshift $z_0$ we derive the values of $\tau_{m,max}$, as a function of the progenitor mass ratio $\xi_{min}$, corresponding to the redshift $z$ for which Eq.~\ref{Eq.NM} gives
$N_m=1$. This means that a merger event with $\xi \geq \xi_{min}$ is still producing a disturbance in all clusters of our sample at the epoch of observation.
%This means that each cluster had a single merger event with $\xi\ge\xi_{min}$ between $z_0$ and the redshift $z$. 
The derived distribution of $\tau_{m,max}$, for each value of $\xi_{min}$, and its mean value are reported in Fig.\ref{Fig.tau_ximin} (blue region and line, respectively). We note that values of $\tau_{m}$ constrained by the observations are always smaller than $\tau_{m,max}$.

%------------------------figure 3----------------------------------------------------
   \begin{figure}
   \centering
\includegraphics[width=7cm]{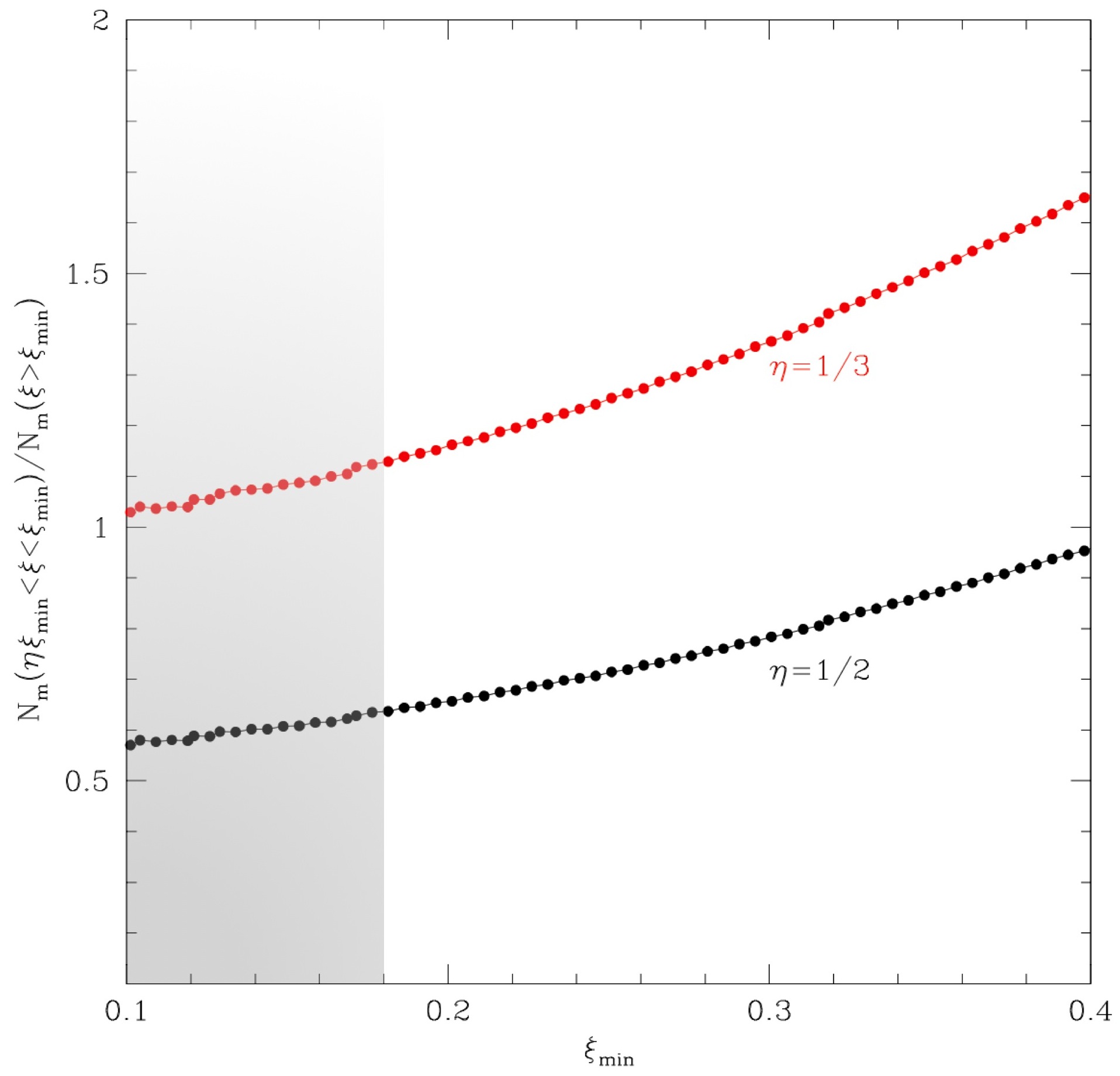}
%{mergerhalo_fraction_35_51.ps}
      \caption{Ratio between the number of mergers with $\eta~\xi_{min}<\xi<\xi_{min}$ and that with $\xi\geq\xi_{min}$, for $\eta=1/2$ (bottom black dots) and $\eta=1/3$ (upper red dots). The shadowed region indicate the range of $\xi_{min}=0.1-0.18$ constrained by the merger timescale derived from numerical simulations (Tormen et al. 2004).}
         \label{Fig.ratioxmin}
   \end{figure}
%-------------------------------------------------------------------------------------

Due to the degeneracy between $\tau_m$ and $\xi_{min}$ in principle large merger timescales can be admitted to explain the observed merger fraction. However, as already anticipated, under this condition our hypothesis of binary mergers can be no longer valid since multiple merger events with mass ratio slightly smaller than $\xi_{min}$ can contribute to the disturbance.
To check this hydrodynamical simulations are necessary to unambiguously relate the merger mass ratio to the cluster morphological parameters, but this is beyond the scope of the paper and deserves future {\it ad hoc} simulations. Here we limit at the following test. 
In Fig.\ref{Fig.ratioxmin} we use the values ($\tau_m$, $\xi_{min}$) constrained in Fig.\ref{Fig.tau_ximin} and show the ratio between the number of mergers with mass ratio in the range $\eta~\xi_{min}-\xi_{min}$ and that of mergers with $\xi\geq\xi_{min}$; $\eta=1/2$ and $1/3$ are considered. We conclude that the binary approach adopted in our paper is appropriate for merger timescale as large as $3-4$ Gyr (those corresponding to $\xi_{min}\sim 0.2-0.3$).

%already shows that there are not many mergers with $\xi$ similar to $\xi_{min}$, at least up to $\xi_{min}\sim 0.2-0.3$, in the time $\tau_m$ constrained in Fig.~\ref{Fig.tau_ximin}.
A possibility to break the degeneracy between $\tau_m$ and $\xi_{min}$ is to adopt values of $\tau_m$ inferred from numerical simulations. A reference timescale is the time necessary to a subcluster to complete an orbit around the center of mass of the main cluster.
%the results of numerical simulations to compute the merger timescale as the time necessary to a subcluster to complete an orbit  around the center of mass of the main cluster. 
Following Tormen et al. (2004) this time can be estimated as $2\times (\tau_a-\tau_p)$, where $\tau_a=\tau_a(\xi)=1.6(\xi+0.02)^{-0.17}$ Gyr and $\tau_p=0.9$ Gyr are the apocentric and pericentric timescale, respectively.  This timescale is reported in Fig.~\ref{Fig.tau_ximin} (green line) as a function of $\xi_{min}$. It intercepts the region constrained by the observations for $\tau_m\simeq 2.5$ Gyr implying $\xi_{min}\simeq 0.14$. In Tormen et al. (2004), $\tau_a$ and $\tau_p$ are derived from analytic fits to the results of numerical simulations, we note however that the dispersion around the median value is $\sim 0.5$ Gyr. As a consequence the merger timescale is constrained as $\tau_m\sim2-3$ Gyr, implying $\xi_{min}\sim 0.1-0.18$.
%he best fit value is 2.5 Gyr, however we consider that the scatter around this timescale can be as large as 0.5 Gyr; \eg Tormen et al. 2004}.
Coming back to Fig.\ref{Fig.ratioxmin} we note that for $\xi_{min}\simeq 0.1-0.18$ (shadowed region) the number of mergers with mass ratio $1/3~\xi_{min}<\xi<\xi_{min}$ is about the same of that of mergers with $\xi\geq\xi_{min}$ and that the number of mergers with $1/2~\xi_{min}<\xi<\xi_{min}$ is about half of that of mergers with $\xi\geq\xi_{min}$. These mergers being characterized by relatively small mass ratio ($\xi\sim0.03-0.09$) and being not numerous should have a negligible influence on the cluster morphological parameters, thus our assumption of binary mergers is reasonably correct.

\subsection{Mergers and Radio Halos}

RH are always observed in dynamically disturbed systems. However they are found only in a fraction of the clusters that are classified as merging systems. In this Section we follow the approach already adopted in Sect.~4.2 and attempt to constrain the properties that cluster mergers should have in order to explain the observed fraction of clusters with RH. 
In Fig.~\ref{Fig.tau_ximin} we report the region ($\tau_m$, $\xi_{min}$) constrained by requiring that the predicted merger fraction matches the observed fraction of clusters with RH (red shadowed region). 
At this point we can adopt two main scenarios: 

\noindent {\it i)} we can assume that the lifetime of RH is equivalent to the lifetime of the merger-induced disturbances identified by X-ray parameters. Under this hypothesis and since RH are found in disturbed systems, we can assume that RH are generated in those systems that have larger mass ratios among merging clusters in our sample. In this case we derive $\xi_{min} \sim 0.18-0.25$ for $\tau_m=2-3$ Gyr as constrained in Sect.4.3.

\noindent {\it ii)} we can assume that RH are statistically generated in ``all'' mergers identified in our sample, but they are short-living compared to the timescale of the merger-induced disturbance. Under this assumption the lifetime of RH is simply $\tau_{RH}\sim \tau_m \times (f_{RH}/f_m)\sim (0.7-0.8)\tau_m$, \ie $\tau_{RH}\sim 1.4-2.4$ Gyr for $\tau_m=2-3$ Gyr.

\subsection{Constraints on $\xi_{min}$ from observations}

The mass ratio $\xi_{min}$ is a simple outcome of our procedure that is appropriate for binary mergers. However this value can be constrained independently by observations of clusters in our sample. 
Observations of single clusters may be used to derive independent constraints on $\xi$.
We collected information in the literature about the mass ratio of the merging clusters of our sample. These mass ratio are derived from optical/near-IR observations of galaxy clusters through the weak lensing analysis or through the study of the galaxy velocity distriubutions in the clusters (see Tab.~\ref{Tab.A1}).
%of optical/near-IR observations of galaxy clusters and fromthe study of the velocity distribution of galaxy in the clusters
%optical and near-IR observations of galaxy clusters in the sample that constrain $\xi_{min}\simeq 0.14-0.16$ through weak lensing analysis or study of the velocity distribution of galaxies in the clusters. 
%Most of the information comes from weak lensing analysis and/or optical observations of galaxy clusters (see Tab.~\ref{Tab.A1}). %and Fig.\ref{Fig.massratio}).
We found information for 7 clusters with RH and for Z5247 that hosts a ``candidate'' RH (see Tab.~\ref{Tab.A1}).
%: Abell 773 (multiple merger with mass ratio 1:4-1:10; Barrena et al. 2007); Abell 1351 (1:5, taking the average mass values for the two substructures; Barrena et al. 2014); Abell 2744 (1:3; Boschin et al. 2006);  Abell 2163 (1:3; Soucail 2012); Abell 1758a (1:2; Okabe \& Umetsu 2008); Abell 520 (1:1; Mahdavi et al. 2007); Abell 1300 (1:1; Ziparo et al. 2012).
For 5 merging clusters without RH, we have looked at the reconstructed convergence maps from weak gravitational lensing (see Tab.~\ref{Tab.A1}). For these clusters we have estimated the mass ratio of the different merging components adopting circular filters on the reconstructed convergence maps with a typical scale that allows to isolate the different correlated peaks. Due to the restricted number density of sources beyond the clusters from which the weak lensing signal is measured and so to the limited resolution of the recovered convergence maps -- around the arcmin scale -- we stress that in these cases the quoted values represent an upper limit for the mass ratio. 
%We derived these estimates for: Abell 2631 (1:7-1:6; Okabe et al 10); Abell 68 (1:6-1:5; Okabe et al 10); A1763 (1:5-1:4; Bardeau et al. 07); Abell 1576 (1:4-1:3; Dahle et al. 02); Abell 781 (1:3; Wittman et al. 14). 
%We also found information for Z5247 (1:4; Dahle et al. 02), that hosts a ``candidate'' RH.

The minimum values of $\xi$ we found for merging clusters in our sample is $\sim0.14-0.16$.
%, these values refer to merging clusters without RH, for which values of $\xi$ as large as $\sim$0.25-0.3 are also measured. 
If we use these values, $\xi_{min}\sim 0.14-0.16$ (black rectangular region in Fig.~\ref{Fig.tau_ximin}), we can derive $\tau_m\simeq 2.5-3$ Gyr, consistent with the reference values of merging timescales derived from 
results of cosmological simulations (see Fig~.\ref{Fig.tau_ximin}).  
 
We found some evidence (the information is available only for half of the merging clusters in the sample) that merging clusters with RH are in general characterized by mergers with larger mass ratio than merging clusters without RH: $\xi$ ranges from $\xi\sim$0.2-0.25 up to $\xi\sim$1. If we assume $\xi_{min}\sim 0.2-0.25$ for merging clusters with RH (red rectangular region in Fig.~\ref{Fig.tau_ximin}), we constrain $\tau_{RH}\sim\tau_m\sim 2.5-3$ Gyr, thus in this case the RH lifetime would be comparable with the merger timescale, potentially supporting the scenario {\it i)} in Sect.4.3.

\section{Summary and Discussion}

Observations establish a clear connection between RH and mergers (\eg Cassano et al. 2010), suggesting that RH can be used as probes of cluster merging rate with cosmic time.
Based on this possibility, in this paper we carry out an exploratory study. By combining the observed fraction of merging clusters and the observed fraction of RH in clusters with the merging rate predicted by fitting formulae based on cosmological simulations we attempt to infer constraints on merger properties of clusters that appear disturbed in X-rays and of clusters that host RH.

We use the Planck SZ cluster catalogue (PSZ; Planck Collaboration XXIX 2014a) and select a sample of 54 clusters with mass $M_{500}\gtrsim 6\times10^{14} \,M_\odot$ and redshift  $0.2\leq z\leq0.33$. 39 of these clusters have both X-ray and radio information and represent a sub-sample that can be used to measure the fraction of RH and that of merging clusters. Mergers in the sample are identified by means of X-ray morphological parameters. 
We find that all RH are in merging clusters whereas not all merging clusters host a RH, specifically $\sim44-51\%$ of the clusters in the sample have a RH, while (using Chandra and XMM-Newton X-ray data) the total fraction of merging clusters is $\sim62-67\%$. 

We convert the theoretical merging rate per halo (FMB10), that mainly depends on the mass ratio of the two progenitors, $\xi=M_i/M_1<1$, into merger fraction by adopting a merger time scale $\tau_m$ as a free parameter. The predicted fraction of merging clusters has a strong dependence on $\xi_{min}$, \ie the minimum mass ratio of the mergers (larger is $\xi_{min}$ smaller is the number of mergers; Fig.~\ref{Fig.mergerfraction}) and on $\tau_m$ (larger is $\tau_m$ larger is the expected fraction of merging clusters; Fig.~\ref{Fig.mergerfraction}).
The comparison between the observed and predicted merger fraction allows to constrain a allowed region in the diagram ($\xi_{min}$, $\tau_m$) where there is degeneracy between these two parameters (Fig~.\ref{Fig.tau_ximin}).

\noindent We attempt to break the degeneracy between $\xi_{min}$ and $\tau_m$: 

\noindent {\it a)} by assuming the merger timescale that is derived by cosmological simulations (\eg Tormen et al. 2004), $\tau_m\sim 2-3$ Gyr, we find that a value $\xi_{min}\sim 0.1-0.18$ explains the observed merger fraction;

\noindent {\it b)} by assuming values of $\xi_{min}$ derived through the analysis of optical/near-IR observations of merging clusters in the sample: $\xi_{min}\sim0.14-0.16$ we find that a merger timescale $\tau_m\simeq 2.5-3$ Gyr explains the observed merger fraction. 

\noindent
Interestingly, values of the parameters that are obtained independently in {\it a)} and {\it b)} are consistent.

%these values of $\tau_m$ are consistent with those derived by cosmological simulations (see Fig~.\ref{Fig.tau_ximin}).  

We find that all clusters with RH in our sample are merging systems but that not all merging clusters host a RH. There are two main possibility to interpret this difference: 

\noindent {\bf \it - Scenario 1)} RH have lifetime similar to the lifetime of merger-driven disturbances in the X-rays, but they are generated in the merging events with larger mass ratio. Values of mass ratio derived from optical/near-IR observations of galaxy clusters in the sample (through weak lensing analysis or through the study of the velocity distribution of galaxies in the clusters) may support this possibility. Indeed we find that $\xi $ ranges between $\sim0.2-0.25$ and $\sim 1$ for RH clusters, whereas values of $\xi\sim0.14-0.16$ up to $\xi\sim0.25-0.3$ are found for merging clusters without RH. If we assume $\xi_{min}\sim 0.2-0.25$, we find that $\tau_{RH}\sim \tau_m\simeq 2-2.5$ Gyr should be adopted to explain the observed RH fraction.
%, in other words we can assume that RH have a lifetime comparable with the merger timescale.

%. These values of $\tau_m$ are equal to those constrained from the observed $\xi_{min}$ for the total merger fraction (see {\it b)} and Fig.~\ref{Fig.tau_ximin}) and quite consistent with the merger timescale constrained with cosmological simulations (\eg Tormen et al. 2004).

\noindent  {\bf \it - Scenario 2)} the lifetime of RH ($\tau_{RH}$) is shorter than the time-scale of merger-induced disturbance in the X-rays, with $\tau_{RH}\sim \tau_m \times (f_{RH}/f_m)\sim (0.7-0.8)\tau_m$. In this case assuming no difference between the mass ratio of clusters with and without RH, we find $\tau_{RH}\sim 1.4-2.4$ Gyr. 

\noindent  In general, we note that both different timescales ($\tau_{RH}\ltsim \tau_m$; \ie {\it Scenario 2}) and mass ratios (\ie {\it Scenario 1}) are likely to govern the statistic of giant RH.

This study deals with several limitations and is based on simplified assumptions:

\noindent {\it i)} the observed fraction of merging clusters is derived by measuring the fraction of clusters with significant X-ray disturbance, this means that we are sensitive only to merger episodes with relevant mass accretion;
%, \ie $\xi\gtsim 0.1$;

\noindent {\it ii)} 
while observationally, the fraction of merging clusters is derived by measuring the morphological disturbances of clusters in the sample on a circular region of $\sim 1$ Mpc (diameter), the theoretical merging rate (FBMI10) and hence the merger fraction is derived by considering all the infalling halos within the virial radius ($\sim 2-3$ Mpc) of the main clusters. However, it should be mentioned that numerical simulations allow to argue that for $\xi\gtsim 0.1$ the two rates should be comparable (see Sect.4 and Tormen et al 2004 for more details);

\noindent {\it iii)} we assume that the X-ray disturbance that we measure in the X-rays is mainly caused by a binary merger event, specifically by the one with larger mass ratio. Thus in deriving the expected merger fraction from theoretical fitting formulae we attempt to select the values of $\xi_{min}$ that matches the merger fraction assuming that mergers with smaller mass ratio do not play a role.
%we assume that in a merger timescale there is a main binary merger event with $\xi\gtsim \xi_{min}$. 
However, in principle, in the timescale of the merger-induced disturbance, $\tau_m$, clusters might experience multiple merger episodes 
with slightly lower mass ratio 
%($\eta\xi_{min}<\xi<\xi_{min}$, with $\eta=1/2,1/3$) 
that can contribute to the morphological disturbance. We show that for typical merger timescale (constrained by
simulations and by the observed value of $\xi_{min}$) the contribution of these slightly minor mergers is expected to be not relevant. Clearly, {\it ad hoc} simulations and follow up studies are necessary to establish a more solid connection between mergers and X-ray disturbances.

It is currently thought that giant RH are generated as a consequence of the acceleration of relativistic electrons by the MHD turbulence stirred up in the ICM by cluster-cluster mergers (\eg Brunetti \& Jones 2014). In this framework the {\it scenario 1)} discussed above implies that the timescale of the X-ray merger-induced disturbances and that of the turbulent stirring of the ICM by cluster mergers in the central 1 Mpc (diameter) region are similar. These timescales are shorter than the dynamical time-scale of the merger which is defined as the time interval between the moment when the center of the less massive cluster first crosses the virial radius of the main one and the moment when the final system reaches a relaxed state. As a consequence, in this scenario RH are not switched on at the beginning of the merger but after a time period that is necessary to the infalling subcluster to generate ICM turbulence in the central Mpc region ($\sim0.9$ Gyr; see Sect.4.1).
Since gravity drives mergers between galaxy clusters, it is expected that the turbulent energy budget should scale with the cluster thermal energy. As a consequence, very massive and merging systems should be the natural host of Mpc-scale RH
%, and thus as about $M_v^{5/3}$ 
(\eg Cassano \& Brunetti 2005; Vazza et al. 2006, 2011; Paul et al. 2011). In line with these expectations we find $f_{RH}\simeq 44-51\%$ for clusters with $M_{500}\gtsim 6\times10^{14} M_{\odot}$ at $0.2\ltsim z\ltsim 0.33$ (see also Sommer \& Basu 2014; Cuciti et al. 2015).
Also the mass-ratio may play a role because major mergers are more powerful events and have the potential to generate more turbulence in larger volumes.
For instance, using a semi-analytic approach, Cassano \& Brunetti (2005) showed that, for a given cluster mass, the ratio between the turbulent energy and the cluster thermal energy increases with increasing $\xi$, becoming smaller than 5\% for $\xi<0.2$ (see also Fig.3 in Cassano \& Brunetti 2005).
This can explain the absence of RH in clusters undergoing merger events with mass ratio $\xi<0.2$.

 On the other hand, {\it scenario 2)} would imply that mergers drive turbulent re-acceleration of relativistic particles in the ICM on a timescale that is $\sim 0.7-0.8$ shorter than the timescale duration of the morphological disturbances in the X-rays. Lagrangian (SPH) simulations of two colliding idealized clusters have been used to study the time-evolution of the RH emission during mergers (Donnert et al. 2013). These simulations predict shortly living RH that are generated after the first core passage and fading within $\ltsim1$ Gyr timescale. In fact, this timescale is shorter than that constrained assuming the scenario 2), however this can be due to the idealized setup of the model and it is very likely that the lifetime of RH is significantly larger in a cosmological contest.
High resolution cosmological simulations also
%Also numerical simulations
% that follow the time-dependent cascade of hydrodynamic turbulence in the ICM of a massive galaxy cluster show that turbulence is generated in the ICM during episode of both minor and major merger events, however 
show an increase of the turbulence (both compressible and incompressible) and of the acceleration rate during 
%episode of high mass accretion rate or 
major mergers (Miniati 2015).
% an increase of the turbulent acceleration rate that can explain the generation of Mpc-scale synchrotron radio halos (Miniati 2015). 
These simulations add also important information as they allow to evaluate the ratio $\tau_{RH}/\tau_m$ under different assumptions about the ICM microphysics. This is an important point as it implies that statistical studies of the connection between RH and merging rates combined with numerical simulations have also the potential to put fundamental constraints on the ICM microphysics and acceleration mechanisms (Miniati 2015, Brunetti 2016).

%also show that 
%during a major merger (with $\xi=0.4$) 
%the resulting acceleration rate of relativistic particles stands for $\sim 1-2$ Gyr, depending on the details of the microphysics, above the levels required to produce Mpc-scale synchrotron emission at GHz frequencies. 
%This is an important argument to push on further investigation of the statistical connection between the RH and cluster merging rate, since constraints on the radio halo lifetime can be used to bind the microphysical properties of the ICM. 

To conclude, while it is clear that massive and merging clusters are the natural hosts of giant RH, the presence of merging clusters without RH pose fundamental questions: 
is there a role of the merger mass ratio in the formation of giant RH? which is the lifetime of RH with respect to the merger timescale? is the RH lifetime tied by the microphysics of the ICM?

Our exploratory study has shown that meaningful values of the merger parameters can be derived combining the observed fraction of RH and the theoretical merging rate in the LCDM model. More specifically our results seem to suggest that the mass-ratio may play a role in the generation of RH, however this result is not conclusive and we cannot conclude whether scenario 1) is favored with respect to scenario 2), or whether mass ratios and different timescales both play a role. An important step forward to address the lifetime of RH and the connection with mergers can be achieved by increasing the statistic of merging clusters without RH. Our study is limited by current data that allow to infer these constraints only in very massive clusters at relatively low redshifts.
In fact in our study we use the currently most complete mass-selected sample of galaxy clusters with radio and X-ray information that, however, is limited to very massive ($M_{500}\gtsim 6\times 10^{14}\, M_{\odot}$) and relatively nearby systems ($z\simeq 0.2-0.33$). Based on energy arguments the occurrence of ``radio quiet'' merging clusters should increase at smaller masses (or at higher redshifts or small mass-ratio mergers; \eg Cassano et al. 2006).
Thus extending the samples of clusters at smaller masses (or at higher redshift) is necessary to obtain stronger constraints on the physical conditions necessary to generate RH. This will be possible with the upcoming new generation of radio facilities, such as LOFAR, ASKAP, MeerKAT up to SKA1(\eg Cassano et al. 2015).
Future surveys also offer the possibility to explore the frequency dependence of the occurrence of RH in galaxy clusters. Current models predict the presence of RH with ultra-steep radio spectra (USSRH) especially in low massive (or high-z) galaxy clusters (\eg Cassano et al. 2006; Brunetti et al. 2008). As a consequence, it is possible that some of the merging clusters without RH host actually USSRH and observations with LOFAR (and SKA1-LOW in the future) will be crucial to check this possibility.

%- multiple mergers

%- mergers with lower mass ratio, $\xi<0.2$ (or $>$ 1:5), that potentially can generate radio halos only in the most massive systems (\eg the ``Bullet'' cluster), are favored to produce large scale radio relics, since would be characterized by shocks with relatively high Mach numbers ($\sim2-3$; \eg Gabici \& Blasi 2003). There are in the literature examples of clusters with double relics systems and without radio halos: \eg Abell 3376 (Bagchi et al. 2006); Abell 1240 (Bonafede et al. 2009) and ZwCl0008 (van Weeren et al. 2011) and these are all low massive clusters ($M_{500}\ltsim 4\times10^{14}\,M_{\odot}$), have also these mergers lower mass ratio?
%Then there are famous merging clusters with radio relics and central halos, CIZA J2242.8+5301 (the Sausage) and 1RXS J0603.1+4214 (the toothbrush) that are both massive and almost equal mass mergers. 
%For instance, the Sausage cluster has a total mass $M_{200}=(2.51\pm0.53)\times10^{15} M_{\odot}$ and is characterized by a merger with $\xi=0.89$ (or 1: 1.12) (Jee et al. 2015).
%Abell 2146 (Russell et al. 2011);
 
%- USSRH?
 
\begin{acknowledgements} 
We thank the referee for the valuable report. S.E. acknowledges support from ASI-INAF n.I/009/10/0 and NuSTAR-ASI/INAF n. I/037/12/0.
GB and RC acknowledge partial support from PRIN-INAF 2014.
GB acknowledges support from the Alexander von Humboldt Foundation.  
We thank K. Dolag for useful discussion and V. Cuciti for providing the morphological parameters of 5 clusters (marked with $^{***}$ in Tab.A1).
%This research was partially supported
\end{acknowledgements}
%-------------------------------------------------------------------

\appendix 

\section{Additional Tables}

\begin{table*}
\caption[]{Cluster's properties}\label{Tab.A1}
\begin{center}
\footnotesize
%\vspace*{-0.8 cm}
\begin{tabular}{llrrccc}
%\hline\noalign{\smallskip}
cluster name & z & $M_{500}$                    & $M_{vir}$                       & radio info & X-ray info  & mass ratio$^{z}$\\
	               &    & [$10^{14} M_{\odot}$] & [$10^{14} M_{\odot}$] &            &            &                        \\	
%\begin{table}[htbp]
%\begin{center}
%\begin{tabular}{l c c c c c c}
%\hline
%\hline
%cluster name &    RA    &      Dec&         z& $M_{500} ( 10^{14}\,M_\odot$)&  radio info& Xinfo\\
\hline
%\vspace*{0.3 cm}
clusters with radio and X-ray data\\
%\vspace*{0.3 cm}
\hline
%\vspace*{0.3 cm}
A2697 & 0.23   & 6.00& 11.00 & UL$^a$ & relaxed$^{1}$& -- \\ 
A3088 & 0.25 & 6.71  & 12.25 & UL$^a$ & relaxed$^{2}$   & -- \\
A2667 & 0.23 &  6.81  & 12.46 & UL$^a$ & relaxed$^{3}$ & -- \\
RXJ0142.0+21 & 0.28 & 6.07 & 10.95 & UL$^b$ & relaxed$^{2}$  & -- \\
A1423 & 0.21 & 6.09 & 11.08 & UL$^a$ & relaxed$^{3}$ & -- \\
A1576 & 0.30 & 5.98 & 10.8 & UL$^b$ & relaxed$^{2,*,**}$  & 1:4-1:3 (Dahle et al. 2002) \\
A2261 & 0.22 & 7.39 & 13.56 & UL$^b$ & relaxed$^{3}$ & -- \\
A2537 & 0.30 & 6.17 & 11.15 & UL$^a$ & relaxed$^{3}$ & -- \\
S0780 & 0.24 & 7.71 & 14.22 & MH$^c$ & relaxed$^{3}$ & --\\
A1835 & 0.25 & 8.46 & 15.53 & MH$^d$ & relaxed$^{4}$ & --\\
A2390 & 0.23 & 9.48 & 17.59 & MH$^e$  & relaxed$^{3}$ & --\\
RXCJ1504.1-02 & 0.22 & 6.98 & 12.80 & MH$^f$  & relaxed$^{3}$ & --\\
A3444 & 0.25 & 7.62 & 13.98 & MH$^c$ & relaxed$^{4}$ & -- \\
A68 & 0.26 & 6.19 & 11.25 & UL$^c$ & merger$^{4}$ & 1:6-1:5 (Okabe et al. 2010) \\
A2631&         0.28&       6.97& 12.75 & UL$^a$& merger$^{3}$ & 1:7-1:6 (Okabe et al. 2010)\\
A781&       	  0.30&       6.36& 11.50 & UL$^{a,ax}$& merger$^{3}$ & 1:3  (Wittman et al. 2014)\\
A1763&		  0.23&	  	   8.29& 15.25 & no RH$^a$ &merger$^{4}$ & 1:5-1:4 (Bardeau et al. 2007)\\
PSZ1 G205.07-62.94	& 0.31& 			7.37&   13.40   & no RH$^g$ & merger$^{1}$ &--\\
%\hline
A2744&  0.31& 9.56& 17.48 & RH$^h$ & merger$^{3}$ & 1:3 (Boschin et al. 2006)\\
A209&	   0.21&	   8.17& 15.02 & RH$^i$&merger$^{3}$ & -- \\
A2163&   0.20&		   16.44 & 31.52 & RH$^l$& merger$^{3}$ & 1:3 (Soucail 2012) \\
RXCJ2003.5-2323&     0.32&       7.48& 13.57 & RH$^i$& merger$^{3}$ & %major M (Giacintucci et al. 2009)
--\\
A520&     0.20&       7.06& 13.06 & RH$^h$& merger$^{3}$ & 1:1 (Mahdavi et al. 07) \\
A773&     0.22&       7.08& 13.05 & RH$^h$& merger$^{3}$ & 1:4-1:10 (Barrera et al. 2007)\\
A1758a&   0.28&       7.99 & 14.68 & RH$^m$ &merger$^{3}$ & 1:2 (Okabe et al. 2008)\\
A1351&     0.32&       7.14& 12.95 & RH$^n$& merger$^{4}$ & 1:5 (Barrera et al. 2014)\\
A2219&     0.23& 11.01& 20.40  & RH$^e$& merger$^{3}$ & 
%complex (Boschin et al. 2004)
\\
A521&       0.25&       6.91& 12.37 & RH$^o$&merger$^{3}$ & 
%complex (ref...)
\\
A697&	    0.28& 11.48& 21.37 & RH$^a$&merger$^{3}$ & 
%complex l.o.s. M (Girardi et al. 2006) 
\\
PSZ1 G171.96-40.64		&  0.27&		   	11.13& 20.88 &RH$^{p}$&merger$^{1}$&--\\
A1300&  0.31&		   8.83& 16.15 & RH$^q$&merger$^{3}$ &1:1 (Ziparo et al. 2011)\\
RXC J1314.4-2515&    0.24&       6.15& 11.20 & RH$^i$ &merger$^{1}$ &  
%$\sim$ equal mass M (Valtchanov et al. 2002)
\\
RXC J1514.9-1523&    0.22&	   8.34& 15.35 & RH$^f$& merger$^{4}$ & --\\
A1682&         0.23&       6.20& 11.33 & RH$^a$ & merger$^{3}$ &
%bimodal (ref)
\\
A1443 &  0.27  &        7.74	& 14.15 & RH$^s$ & merger$^{***}$ 	&--\\
Z5247&	0.23&		6.04 & 11.00 & RH?$^c$ & merger$^{4}$ & 1:4 (Dahle et al. 2002)\\
A2552&	0.30&		7.53&  13.65 & RH?$^c$ & relaxed?$^{4,*}$ & --\\     
RXC J0510.7-0801& 0.22&	7.36& 13.50 & RH?$^c$  & merger$^{\surd}$ &--\\
A402 & 0.32 & 7.20 & 13.06 & MH?$^r$ & relaxed$^{***}$ & --\\
\hline
clusters with (only) X-ray data\\
\hline
A2895 &	0.23&		6.15& 11.22 & - & merger$^{\surd}$  & --\\
A2813 & 0.29&		9.16& 15.15 & -  & merger$^{\surd}$  & -- \\
PSZ1G139.61+24 & 0.27 & 7.09 & 12.93 & -- & relaxed$^{***}$  & -- \\
%A1443 &  0.27  &        7.74	& 14.15 &-- & M$^{\surd}$ 	&--\\
A2355 &  0.23         & 6.92 & 12.75 & -- & merger$^{***}$  &-- \\
A1733 &  0.26        & 7.05 & 12.85 & -- & merger$^{***}$ &-- \\
MACS J2135-010   		& 0.33      &  7.57 & 13.8 & --& merger$^{\surd}$  & --\\
RXC J2051.1+0216     & 0.32   &     6.13 & 11.07 & -- & merger$^{1}$&--\\
A2472	 & 0.31 & 6.15 & 11.12 & -- & merger$^{1}$ & --\\
A56	  & 0.30 &       6.20& 11.25 & -- &merger$^{1}$&--\\
A384 & 0.24 & 	6.38 & 11.65 & -- & relaxed$^{1}$ &--\\
RXCJ1322.8+31 & 0.31 & 6.63 & 12.0 & -- & relaxed$^{1}$ &--\\
PSZ1 G019.12+3123    &   0.28       &   7.08 & 12.95 & -- & merger$^{1}$ & --\\
\hline
clusters without radio and X-ray data\\
\hline
ZwCl 1028.8+1419$^*$ & 0.31     &         6.11	&11.05  & --& --& --\\ 
A3041$^*$ 	 & 0.23    &          6.12& 11.24 &-- &--& --\\
A220$^*$ 	 & 0.33   &       	   6.74&12.20 & --&--& --\\
%\hline
%\end{tabular}
%\caption[high redshift sample cluster properties]{high redshift sample clusters porperties} \label{tab:highzsample}
%\end{center}
%\end{table} 
\noalign{\smallskip}
\hline\noalign{\smallskip}
\end{tabular}
\end{center}
\vspace*{-0.5 cm}
\footnotesize Notes -- radio info.:$^a$ Venturi et al. 2008; $^b$ Kale et al. 2013; $^c$ Kale et al. 2015; $^d$ Murgia et al. 2009; $^e$ Bacchi et al. 2003; $^f$ Giacintucci et al. 2011; $^g$ Ferrari et al. (private communication); $^h$ Govoni et al. 2001; $^i$ Venturi et al. 2007; $^l$ Feretti et al. 2001; $^m$ Giovannini et al. 2006; $^n$ Giacintucci et al. 2009;$^o$ Brunetti et al. 2008; $^{p}$ Giacintucci et al. 2013; $^q$ Reid et al. 1999; $^r$ Macario et al. (private communication); $^s$ Bonafede et al. 2015. X-ray info.: $^{1}$ from XMM-Newton visual inspection; $^{2}$ Cassano et al. 2013; $^{3}$ Cassano et al 2010; $^{4}$ Cuciti et al. 2015; $^{\surd}$ this paper. 
$^{ax}$ Venturi et al. (2011) report on a possible RH in A781, however those observations were not conclusive (see also Govoni et al. 2011). $^*$ Landry et al. (2013) classify these clusters as ``unrelaxed'' systems; $^{**}$ Dahle et al. (2002) report on some merger activity in this cluster. $^z$ the mass ratio here is defined as $1/\xi$. $^{***}$ V. Cuciti, private communication.
%, based on the analysis of weak lensing data.

%Clusters belonging to the EGRHS are in the upper part of table, while clusters only in the PSZ catalog are in the lower part. $M_{500}$ is in units of $10^{14}\, M_{\odot}$ (from Planck Collaboration 2013a). Radio analysis of clusters belonging to the EGRHS is published in Venturi et al. (2007; 2008); Kale et al. (2013), and references therein. Information about the presence of diffuse radio emission is reported in col.6: RH=radio halo; MH=mini halo; UL=upper limit. X-ray data of EGRHS clusters are analyzed in Cassano et al. (2010, 2013, and reference therein). Analysis of GMRT data is in progress for some clusters observed during the runs 22-029 and 23-004 (see Table). Only poor quality data are available for A\,2895 and A\,2813 (Venturi et al. 2008).
%The symbol $\surd$ indicates the availability of radio and X-ray archival information.
%   $^c$ L-band VLA locked observations in C and D configurations; $^v$Venturi et al. 2008.

 %$^b$ Macario et al. (private communication); $^c$ L-band VLA locked observations in C and D configurations. Clusters which need additional radio observations are in boldface.
%\label{tab:PLCK171}
\end{table*}

\begin{table*}
\caption[]{Cluster's morphological parameters}\label{Tab.B1}
\begin{center}
\footnotesize
%\vspace*{-0.8 cm}
\begin{tabular}{lrrr}
%\hline\noalign{\smallskip}
cluster name & $P_3/P_0$ [min, max] & $w$ [min, max] & $c$ [min, max] \\
	               &    $[10^{-7}]$  & $\,\,\,\,\,[10^{-2}]$         &          \\	
%\begin{table}[htbp]
%\begin{center}
%\begin{tabular}{l c c c c c c}
%\hline
%\hline
%cluster name &    RA    &      Dec&         z& $M_{500} ( 10^{14}\,M_\odot$)&  radio info& Xinfo\\
%\hline
%\vspace*{0.3 cm}
%clusters with radio and X-ray data\\
%\vspace*{0.3 cm}
\hline
%\vspace*{0.3 cm}
%A2697 &  -- & -- & --\\ 
A3088 &  0.833 [0.279, 1.663] & 0.285 [0.220, 0.370]	& 0.339 [0.332, 0.345]  \\
A2667 &  1.395 [0.799, 2.152] & 0.927 [0.800, 1.030] &	 0.407 [0.402, 0.410] \\
RXJ0142.0+21 & 6.625 [3.655, 10.350] & 0.738 [0.650, 0.910] &  0.186 [0.180, 0.191] \\
A1423 & 1.413 [0.656, 3.880] & 0.562 [0.460, 0.760]  &	 0.331 [0.323, 0.342] \\
A1576 &  5.950 [3.661, 11.071] & 1.271 [0.940, 1.590]  & 0.235 [0.226, 0.241]\\
A2261 & 1.026 [0.513, 1.673] & 0.495 [0.430, 0.570]  &	 0.334 [0.330, 0.337] \\
A2537 & 0.351 [0.165, 1.208] & 0.561 [0.460, 0.660] &	 0.278 [0.273, 0.282] \\
S0780 & 0.480 [0.243, 0.801] & 0.827 [0.760, 0.880]	& 0.473 [0.470, 0.476]\\
A1835 & 0.459 [0.317, 0.576] & 0.996 [0.952,  1.032] &  0.487 [0.485, 0.488] \\
A2390 & 0.694 [0.520, 0.933] & 1.171 [1.120, 1.200] &	 0.305 [0.303, 0.306]\\
RXCJ1504.1-02 & 0.148 [0.086, 0.221] & 0.459 [0.430, 0.490]	& 0.624 [0.622, 0.626] \\
A3444 &  0.434  [0.256, 0.650] & 0.745 [0.683,  0.806] & 0.465 [0.461, 0.467] \\
A68 & 3.199 [1.368, 7.026] & 1.004 [0.740, 1.240]  & 0.149	[0.141,	     0.157]\\
A2631&   1.550 [0.647, 5.941] & 1.570 [1.270, 1.920] &	 0.121 [0.114, 0.128]\\
A781&       3.143 [0.711, 11.880] &  6.374 [5.830, 6.770]	& 0.111 [0.103, 0.118]\\
A1763&	1.222  [0.480, 2.509]	 &	1.885 [1.686, 2.039] & 0.139 	[0.135, 0.143]\\
%PSZ1 G205.07-62.94	& --&   --   &--\\
%\hline
A2744&  11.050 [7.995, 14.070] & 2.637 [2.490, 2.760] &	 0.101 [0.098, 0.103] \\
A209&     0.5.185 [0.136, 1.465] & 1.321 [1.150, 1.460] &	 0.176 [0.170, 0.181]  \\
A2163 &    14.850 [13.770, 16.120]  & 5.970 [5.890, 6.020]  & 0.116 [0.115, 0.118] \\
RXCJ2003.5-2323&    4.602 [2.507, 9.255]  & 1.824 [1.440, 1.970]	 & 0.062 [0.059, 0.064] \\
A520&      5.259 [4.779, 5.588]  & 10.050 [10.030, 10.110] &	 0.0976 [0.0971, 0.0983] \\
A773&    1.445 [0.659, 2.705]  & 2.403 [2.220, 2.530] &	 0.184 [0.179, 0.188] \\
A1758a & 2.515 [1.492, 3.697] &  8.217 [8.070, 8.320]	& 0.109 [0.106, 0.111]\\
A1351&   3.506 [1.900, 7.398] & 4.272	[3.872, 4.527] & 0.083	[0.079,    0.088] \\
A2219&     1.681 [1.228, 2.068]  & 2.127 [2.070, 2.190]	& 0.134 [0.133, 0.136] \\
A521&  5.090 [2.981, 7.771] &  2.204 [2.030, 2.470]	& 0.108 [0.104, 0.111]     \\
A697&  1.668 [0.790, 3.919]  & 0.731 [0.580, 0.890]	 &0.153 [0.149, 0.157]	 \\
%PSZ1 G171.96-40.64 &  -- &-- & -- \\
A1300&  6.847 [4.079, 12.880]  &4.442 [4.230, 4.640] &	 0.191 [0.185, 0.197] \\
%RXC J1314.4-2515&    -- &   -- & -- \\
RXC J1514.9-1523&   1.411 [0.491, 2.995] & 1.301	[1.063, 1.429] & 0.064	[0.062,    0.066]\\
A1682&        15.320 [8.342, 24.490] &  2.054 [1.820, 2.390]	& 0.126 [0.119, 0.132] \\
A1443 &  12.890	[6.644,  	21.020] & 3.530	[3.138, 3.817] & 0.109	[0.101,    	0.115] \\
Z5247&	3.061  [0.744,  8.739]&	3.362	[2.890,  3.667] &0.158	[0.138,     0.173]	\\
A2552&	0.222  [0.106,  1.383] &		0.639 	[0.523,   0.824]  & 0.218 [0.212, 0.224] \\     
RXC J0510.7-0801& 2.171	[0.885, 4.356]	 &  2.346 	[2.140, 2.590] & 0.134	[0.129, 0.138] \\
A402 & 1.350	[0.793,  	3.169] &		1.249	[1.109, 1.399]	 & 0.323 [0.315, 0.331]		  \\
%\hline
%clusters with (only) X-ray data\\
%hline
A2895 &	4.851 [2.929,	 7.732] &			4.271 	[4.020, 4.440] & 0.161 	[0.155,		0.167]\\
A2813 & 1.230	[0.396, 	3.239] &			0.311 	[0.300, 0.550]  & 0.172 	[0.168,		0.176]\\
PSZ1G139.61+24 & 0.194 [0.074, 0.965] &	1.348 	[1.175, 1.526]	& 0.362 [0.358,     	0.369]	  \\
%A1443 &  0.27  &        7.74	& 14.15 &-- & M$^{\surd}$ 	&--\\
A2355  & 7.495	[4.403, 12.331]  &	4.879	 [4.458, 5.066] & 0.075	 [0.071,    	0.080]\\
A1733 &  0.299	[0.629, 7.244] &	4.219 	[3.738, 4.642] & 0.134	[0.121,     	0.142] \\
MACS J2135-010   	&	4.073 [2.203, 10.711] &		1.188 	[0.917, 1.445]& 0.139 	[0.133,    	0.144]\\
%RXC J2051.1+0216     & --  &     --  & -- \\
%A2472	 & -- & --  & --\\
%A56	 & -- & --  & --\\
%A384 & -- & --  & --\\
%RXCJ1322.8+31 & -- & --  & --\\
%PSZ1 G019.12+3123    & -- & --  & --\\
%\hline
%clusters without radio and X-ray data\\
%\hline
%ZwCl 1028.8+1419$^*$ & 0.31     &         6.11	&11.05  & --& --& --\\ 
%A3041$^*$ 	 & 0.23    &          6.12& 11.24 &-- &--& --\\
%A220$^*$ 	 & 0.33   &       	   6.74&12.20 & --&--& --\\
%\hline
%\end{tabular}
%\caption[high redshift sample cluster properties]{high redshift sample clusters porperties} \label{tab:highzsample}
%\end{center}
%\end{table} 
\noalign{\smallskip}
\hline\noalign{\smallskip}
\end{tabular}
\end{center}
\vspace*{-0.5 cm}
\end{table*}

%\section{Additional Tables}
In Tab.\ref{Tab.A1} we report the main properties of the 54 clusters belonging to our sample, specifically: Col.(1) cluster's name; Col.(2) cluster's redshift; Col.(3) $M_{500}$ from Planck Collaboration 2014; Col.(4) virial mass, $M_{vir}$ (see Sect.~2); Col.(5) information about the presence of diffuse radio emission; Col.(6) cluster dynamical status; Col.(7) mass ratio, when available. The first two panels contain the 39 clusters with both radio and X-ray information (clusters with RH are in the second panel); the third panel contains clusters with X-ray information; the three clusters in the forth panel are those without radio and X-ray information.  In Tab.\ref{Tab.B1} we report the morphological parameters derived for the 41 clusters with Chandra X-ray data (see Sect.~2): $P_3/P_0$, $w$ and $c$, with their inferior and superior values ([$P_3/P_0-1\sigma$, $P_3/P_0+1\sigma$], and so on for the others).

%In Fig.\ref{Fig.massratio} we report the distribution of the mass-ratio for merging clusters with RH (red points) and without RH (blue points) collected from the literature.

%------------------------figure 3----------------------------------------------------
%   \begin{figure}
%   \centering
%\includegraphics[width=7cm]{massratio_mix_pap.jpg}
%{mergerhalo_fraction_35_51.ps}
%      \caption{Mass ratio of the merging clusters in the sample with RH (red points) and without RH (blue points) collected from the literature. The blue and red vertical bands show the minimum value of $\xi_{min}$ constrained by these observations for the merging clusters without RH and for the RH clusters, respectively. We note that the cluster A781 could host a RH with a very steep radio spectrum, but additional observations are necessary to firmly detect it (Venturi et al. 2011). }
 %              \label{Fig.massratio}
 %  \end{figure}
%-------------------------------------------------------------------------------------


\begin{thebibliography}{}
\bibitem[Bacchi et 
al.(2003)]{2003A&A...400..465B} Bacchi, M., Feretti, L., Giovannini, G., \& Govoni, F.\ 2003, \aap, 400, 465 
\bibitem[Bardeau et 
al.(2007)]{2007A&A...470..449B} Bardeau, S., Soucail, G., Kneib, J.-P., et al.\ 2007, \aap, 470, 449 

\bibitem[Barrena et al.(2007)]{2007A&A...467...37B} Barrena, R., Boschin, W., Girardi, M., \& Spolaor, M.\ 2007, \aap, 467, 37 
\bibitem[Barrena et al.(2014)]{2014MNRAS.442.2216B} Barrena, R., Girardi, 
M., Boschin, W., De Grandi, S., \& Rossetti, M.\ 2014, \mnras, 442, 2216 

\bibitem {}Basu, K. 2012 MNRAS, 421L, 112

%\bibitem[Bertone \& Conselice(2009)]{2009MNRAS.396.2345B} Bertone, S., \& Conselice, C.~J.\ 2009, \mnras, 396, 2345 

\bibitem{}B\"ohringer, H.; Pratt, G. W.; Arnaud, M.; et al., 2010, A\&A, 514, 32
\bibitem[Bonafede et al.(2015)]{2015MNRAS.454.3391B} Bonafede, A., Intema, 
H., Br{\"u}ggen, M., et al.\ 2015, \mnras, 454, 3391 

\bibitem[Boschin et al.(2006)]{2006A&A...449..461B} Boschin, W., Girardi, M., Spolaor, M., \& Barrena, R.\ 2006, \aap, 449, 461 

\bibitem[Boylan-Kolchin et al.(2009)]{2009MNRAS.398.1150B} Boylan-Kolchin, 
M., Springel, V., White, S.~D.~M., Jenkins, A., 
\& Lemson, G.\ 2009, \mnras, 398, 1150


\bibitem[Br{\"u}ggen et al.(2012)]{2012SSRv..166..187B} Br{\"u}ggen, M., Bykov, A., Ryu, D., R{\"o}ttgering, H.\ 2012, \ssr, 166, 187 
\bibitem[Br{\"u}ggen \& Vazza(2015)]{2015ASSL..407..599B} Br{\"u}ggen, M., \& Vazza, F.\ 2015, Magnetic Fields in Diffuse Media, 407, 599 
  
 \bibitem[Brunetti(2016)]{2016PPCF...58a4011B} Brunetti, G.\ 2016, Plasma 
Physics and Controlled Fusion, 58, 014011  
\bibitem[2007] {B07}Brunetti G., Venturi, T., Dallacasa, D., Cassano, R., Dolag K., Giacintucci, S., Setti, G., 2007, ApJ Letter, 670, L5
\bibitem[Brunetti \& Lazarian(2007)]{2007MNRAS.378..245B} Brunetti, G., \& Lazarian, A.\ 2007, \mnras, 378, 245 
\bibitem[2008]{B08} Brunetti, G., Giacintucci, S., Cassano, R. et al. 2008, Nature, 455, 944
 \bibitem[2009] {B09}Brunetti G., Cassano, R. Dolag, K., Setti, G. 2009, A\&A, 507, 661
\bibitem[Brunetti \& Jones(2014)]{2014IJMPD..2330007B} Brunetti, G., \& Jones, T.~W.\ 2014, International Journal of Modern Physics D, 23, 1430007 

%\bibitem[Buote(2001)]{2001ApJ...553L..15B} Buote, D.~A.\ 2001, \apjl, 553, L15
\bibitem[1995]{Buote95}Buote, D. A. \& Tsai, J. C. 1995, ApJ, 452, 522


\bibitem[Cassano \& Brunetti(2005)]{2005MNRAS.357.1313C} Cassano, R., \& Brunetti, G.\ 2005, \mnras, 357, 1313 
\bibitem[Cassano et al.(2006)]{2006MNRAS.369.1577C} Cassano, R., Brunetti, 
G., \& Setti, G.\ 2006, \mnras, 369, 1577
\bibitem[Cassano et al.(2010)]{2010ApJ...721L..82C} Cassano, R., Ettori, S., Giacintucci, S., et al.\ 2010, \apjl, 721, L82
\bibitem[2013]{C13}Cassano, R., Ettori, S., Brunetti, G., Giacintucci, S., Pratt, G. W. et al. 2013, ApJ, 777, 141
\bibitem[Cassano et al.(2015)]{2015aska.confE..73C} Cassano, R., Bernardi, G., Brunetti, G., et al.\ 2015, Advancing Astrophysics with the Square Kilometre Array (AASKA14), 73 

\bibitem[Conselice(2014)]{2014ARA&A..52..291C} Conselice, C.~J.\ 2014, \araa, 52, 291
\bibitem[Conselice et al.(2003)]{2003AJ....126.1183C} Conselice, C.~J., Bershady, M.~A., Dickinson, M., \& Papovich, C.\ 2003, \aj, 126, 1183 
%\bibitem[Conselice et al.(2008)]{2008MNRAS.386..909C} Conselice, C.~J., Rajgor, S., \& Myers, R.\ 2008, \mnras, 386, 909 
%\bibitem[Conselice et al.(2014)]{2014MNRAS.444.1125C} Conselice, C.~J., Bluck, A.~F.~L., Mortlock, A., Palamara, D., \& Benson, A.~J.\ 2014, \mnras, 444, 1125 

\bibitem[Cuciti et al.(2015)]{2015A&A...580A..97C} Cuciti, V., Cassano, R., Brunetti, G., et al.\ 2015, \aap, 580, A97 


\bibitem[Dahle et al.(2002)]{2002ApJS..139..313D} Dahle, H., Kaiser, N., Irgens, R.~J., Lilje, P.~B., \& Maddox, S.~J.\ 2002, \apjs, 139, 313 

\bibitem[De Boni et al.(2016)]{2015arXiv150101977D} De Boni, C., Serra, A.~L., Diaferio, A., Giocoli, C., \& Baldi, M.\ 2016, ApJ 818, 188 
%arXiv:1501.01977
\bibitem[De Propris et al.(2005)]{2005AJ....130.1516D} De Propris, R., Liske, J., Driver, S.~P., Allen, P.~D., 
\& Cross, N.~J.~G.\ 2005, \aj, 130, 1516 

\bibitem[Diaferio(2015)]{2015arXiv150201195D} Diaferio, A.\ 2015, arXiv:1502.01195 
%\bibitem[Dolag et al.(2005)]{2005MNRAS.364..753D} Dolag, K., Vazza, F., Brunetti, G., \& Tormen, G.\ 2005, \mnras, 364, 753 
\bibitem[Donnert et al.(2013)]{2013MNRAS.429.3564D} Donnert, J., Dolag, K., Brunetti, G., \& Cassano, R.\ 2013, \mnras, 429, 3564 
\bibitem[Duffy et al.(2008)]{2008MNRAS.390L..64D} Duffy, A.~R., Schaye, J., 
Kay, S.~T., \& Dalla Vecchia, C.\ 2008, \mnras, 390, L64 

\bibitem[Ettori et 
al.(2010)]{2010A&A...524A..68E} Ettori, S., Gastaldello, F., Leccardi, A., et al.\ 2010, \aap, 524, A68 



\bibitem[Fakhouri \& Ma(2008)]{2008MNRAS.386..577F} Fakhouri, O., \& Ma, C.-P.\ 2008, \mnras, 386, 577
\bibitem[Fakhouri et al.(2010)]{2010MNRAS.406.2267F} Fakhouri, O., Ma, 
C.-P., \& Boylan-Kolchin, M.\ 2010, \mnras, 406, 2267

\bibitem[Feretti et 
al.(2001)]{2001A&A...373..106F} Feretti, L., Fusco-Femiano, R., Giovannini, G., \& Govoni, F.\ 2001, \aap, 373, 106 

\bibitem{}Feretti, L., Giovannini, G., Govoni, F., Murgia, M. 2012, A\&ARv, 20, 54

\bibitem[Giacintucci et al.(2009)]{2009ApJ...704L..54G} Giacintucci, S., Venturi, T., Cassano, R., Dallacasa, D., \& Brunetti, G.\ 2009, \apjl, 704, L54 
%\bibitem[Giacintucci et al.(2009)]{2009A&A...505...45G} Giacintucci, S., Venturi, T., Brunetti, G., et al.\ 2009, \aap, 505, 45 
\bibitem[Giacintucci et al.(2011)]{2011A&A...534A..57G} Giacintucci, S., Dallacasa, D., Venturi, T., et al.\ 2011, \aap, 534, A57
\bibitem[Giacintucci et al.(2013)]{2013arXiv1302.0218G} Giacintucci, S., Kale, R., Wik, D.~R., Venturi, T., \& Markevitch, M.\ 2013, ApJ, 766, 18


\bibitem[Giovannini et al.(2006)]{2006AN....327..563G} Giovannini, G., 
Feretti, L., Govoni, F., Murgia, M., 
\& Pizzo, R.\ 2006, Astronomische Nachrichten, 327, 563 

%\bibitem[Giovannini et 
%al.(2009)]{2009A&A...507.1257G} Giovannini, G., Bonafede, A., Feretti, L., et al.\ 2009, \aap, 507, 1257 


\bibitem[Giocoli et al.(2007)]{2007MNRAS.376..977G} Giocoli, C., Moreno, 
J., Sheth, R.~K., \& Tormen, G.\ 2007, \mnras, 376, 977 
\bibitem[Giocoli et al.(2012)]{2012MNRAS.422..185G} Giocoli, C., Tormen, G., \& Sheth, R.~K.\ 2012, \mnras, 422, 185

\bibitem{}Govoni, F., En\ss lin, T. A., Feretti, L., Giovannini, G., 2001, A\&A 369, 441
\bibitem[Govoni et 
al.(2011)]{2011A&A...529A..69G} Govoni, F., Murgia, M., Giovannini, G., Vacca, V., \& Bonafede, A.\ 2011, \aap, 529, A69 


\bibitem[Guo \& White(2008)]{2008MNRAS.384....2G} Guo, Q., \& White, S.~D.~M.\ 2008, \mnras, 384, 2 

%\bibitem[Hallman \& Jeltema(2011)]{2011MNRAS.418.2467H} Hallman, E.~J., \& Jeltema, T.~E.\ 2011, \mnras, 418, 2467 
\bibitem[Hopkins et al.(2013)]{2013MNRAS.430.1901H} Hopkins, P.~F., Cox, 
T.~J., Hernquist, L., et al.\ 2013, \mnras, 430, 1901

\bibitem[Jian et al.(2012)]{2012ApJ...754...26J} Jian, H.-Y., Lin, L., 
\& Chiueh, T.\ 2012, \apj, 754, 26 

\bibitem[Kale \& Parekh(2016)]{2016MNRAS.tmp..579K} Kale, R., \& Parekh, V.\ 2016, \mnras,  
\bibitem[Kale et 
al.(2013)]{2013A&A...557A..99K} Kale, R., Venturi, T., Giacintucci, S., et al.\ 2013, \aap, 557, A99
\bibitem[Kale et 
al.(2015)]{2015A&A...579A..92K} Kale, R., Venturi, T., Giacintucci, S., et al.\ 2015, \aap, 579, A92 

\bibitem[Kravtsov 
\& Borgani(2012)]{2012ARA&A..50..353K} Kravtsov, A.~V., \& Borgani, S.\ 2012, \araa, 50, 353 
\bibitem[Kulsrud et al.(1997)]{1997ApJ...480..481K} Kulsrud, R.~M., Cen, 
R., Ostriker, J.~P., \& Ryu, D.\ 1997, \apj, 480, 481 

\bibitem[Landry et al.(2013)]{2013MNRAS.433.2790L} Landry, D., Bonamente, 
M., Giles, P., et al.\ 2013, \mnras, 433, 2790 

\bibitem[Lemze et al.(2013)]{2013ApJ...776...91L} Lemze, D., Postman, M., 
Genel, S., et al.\ 2013, \apj, 776, 91 

\bibitem[Lotz et al.(2004)]{2004AJ....128..163L} Lotz, J.~M., Primack, J., 
\& Madau, P.\ 2004, \aj, 128, 163 
\bibitem[Lotz et al.(2011)]{2011ApJ...742..103L} Lotz, J.~M., Jonsson, P., 
Cox, T.~J., et al.\ 2011, \apj, 742, 103 

\bibitem[Mahdavi et al.(2007)]{2007ApJ...668..806M} Mahdavi, A., Hoekstra, 
H., Babul, A., Balam, D.~D., \& Capak, P.~L.\ 2007, \apj, 668, 806 
\bibitem[Mantz et al.(2015)]{2015MNRAS.449..199M} Mantz, A.~B., Allen, 
S.~W., Morris, R.~G., et al.\ 2015, \mnras, 449, 199 
\bibitem[Miniati(2015)]{2015ApJ...800...60M} Miniati, F.\ 2015, \apj, 800, 
60 
\bibitem[1993]{Mohr93}Mohr, J. J., Fabricant, D. G., Geller, M. J. 1993, ApJ, 413, 492


\bibitem[McBride et al.(2009)]{2009MNRAS.398.1858M} McBride, J., Fakhouri, 
O., \& Ma, C.-P.\ 2009, \mnras, 398, 1858 

\bibitem[Moreno et al.(2008)]{2008MNRAS.391.1729M} Moreno, J., Giocoli, C., 
\& Sheth, R.~K.\ 2008, \mnras, 391, 1729 
\bibitem[Murgia et 
al.(2009)]{2009A&A...499..679M} Murgia, M., Govoni, F., Markevitch, M., et al.\ 2009, \aap, 499, 679 


\bibitem[Navarro et al.(1997)]{1997ApJ...490..493N} Navarro, J.~F., Frenk, 
C.~S., \& White, S.~D.~M.\ 1997, \apj, 490, 493 
\bibitem[Norman 
\& Bryan(1999)]{1999LNP...530..106N} Norman, M.~L., \& Bryan, G.~L.\ 1999, The Radio Galaxy Messier 87, 530, 106 

\bibitem[Okabe \& Umetsu(2008)]{2008PASJ...60..345O} Okabe, N., \& Umetsu, K.\ 2008, \pasj, 60, 345 
\bibitem[Okabe et al.(2010)]{2010PASJ...62..811O} Okabe, N., Takada, M., 
Umetsu, K., Futamase, T., \& Smith, G.~P.\ 2010, \pasj, 62, 811 


\bibitem[Parekh et 
al.(2015)]{2015A&A...575A.127P} Parekh, V., van der Heyden, K., Ferrari, C., Angus, G., \& Holwerda, B.\ 2015, \aap, 575, A127 
\bibitem[Patton et al.(2000)]{2000ApJ...536..153P} Patton, D.~R., Carlberg, 
R.~G., Marzke, R.~O., et al.\ 2000, \apj, 536, 153 
\bibitem[Paul et al.(2011)]{2011ApJ...726...17P} Paul, S., Iapichino, L., Miniati, F., Bagchi, J., \& Mannheim, K.\ 2011, \apj, 726, 17 

\bibitem[Planck Collaboration et 
al.(2014)]{2014A&A...571A..29P} Planck Collaboration, Ade, P.~A.~R., Aghanim, N., et al.\ 2014a, \aap, 571, A29 
\bibitem[Planck Collaboration et 
al.(2014)]{2014A&A...571A..20P} Planck Collaboration, Ade, P.~A.~R., Aghanim, N., et al.\ 2014b, \aap, 571, A20 

\bibitem[Poole et al.(2006)]{2006MNRAS.373..881P} Poole, G.~B., Fardal, 
M.~A., Babul, A., et al.\ 2006, \mnras, 373, 881 


\bibitem[Reid et al.(1999)]{1999MNRAS.302..571R} Reid, A.~D., Hunstead, 
R.~W., Lemonon, L., \& Pierre, M.~M.\ 1999, \mnras, 302, 571 
\bibitem[Ricker \& Sarazin(2001)]{2001ApJ...561..621R} Ricker, P.~M., \& Sarazin, C.~L.\ 2001, \apj, 561, 621
\bibitem[Rossetti et 
al.(2011)]{2011A&A...532A.123R} Rossetti, M., Eckert, D., Cavalleri, B.~M., et al.\ 2011, \aap, 532, A123 

\bibitem[2008]{Santos08}Santos, J.S., Rosati, P., Tozzi, P., B{\"o}hringer, H., Ettori, S., Bignamini, A. 2008, A\&A, 483, 35
%\bibitem[Sarazin(2002)]{2002ASSL..272....1S} Sarazin, C.~L.\ 2002, Merging Processes in Galaxy Clusters, 272, 1 

\bibitem[2014]{SB14}Sommer, M. W., Basu, K. 2014, MNRAS, 437, 2163
\bibitem[Soucail(2012)]{2012A&A...540A..61S} Soucail, G.\ 2012, \aap, 540, A61 
\bibitem[Springel et al.(2005)]{2005Natur.435..629S} Springel, V., White, 
S.~D.~M., Jenkins, A., et al.\ 2005, \nat, 435, 629 
\bibitem[Subramanian et al.(2006)]{2006MNRAS.366.1437S} Subramanian, K., 
Shukurov, A., \& Haugen, N.~E.~L.\ 2006, \mnras, 366, 1437 

\bibitem[Tormen et al.(2004)]{2004MNRAS.350.1397T} Tormen, G., Moscardini, 
L., \& Yoshida, N.\ 2004, \mnras, 350, 1397 

\bibitem[Vazza et al.(2006)]{2006MNRAS.369L..14V} Vazza, F., Tormen, G., 
Cassano, R., Brunetti, G., \& Dolag, K.\ 2006, \mnras, 369, L14 
\bibitem[Vazza et 
al.(2011)]{2011A&A...529A..17V} Vazza, F., Brunetti, G., Gheller, C., Brunino, R., \& Br{\"u}ggen, M.\ 2011, \aap, 529, A17 

\bibitem[Venturi et 
al.(2007)]{2007A&A...463..937V} Venturi, T., Giacintucci, S., Brunetti, G., et al.\ 2007, \aap, 463, 937 
\bibitem[2008]{V08}Venturi, T., Giacintucci, S., Dallacasa, D., Cassano, R., Brunetti, G. et al. 2008, A\&A, 484, 327 
\bibitem[Venturi et al.(2011)]{2011MNRAS.414L..65V} Venturi, T., 
Giacintucci, G., Dallacasa, D., et al.\ 2011, \mnras, 414, L65
\bibitem[van den Bosch(2002)]{2002MNRAS.331...98V} van den Bosch, F.~C.\ 
2002, \mnras, 331, 98

\bibitem[Wen \& Han(2015)]{2015MNRAS.448....2W} Wen, Z.~L., \& Han, J.~L.\ 2015, \mnras, 448, 2
\bibitem[Yuan et al.(2015)]{2015ApJ...813...77Y} Yuan, Z.~S., Han, J.~L., 
\& Wen, Z.~L.\ 2015, \apj, 813, 77 


%  \bibitem[1966]{baker} Baker, N. 1966,
%      in Stellar Evolution,
 %     ed.\ R. F. Stein,\& A. G. W. Cameron
   %   (Plenum, New York) 333

   %\bibitem[1988]{balluch} Balluch, M. 1988  A\&A, 200, 58

\bibitem[Wittman et al.(2014)]{2014MNRAS.437.3578W} Wittman, D., Dawson, 
W., \& Benson, B.\ 2014, \mnras, 437, 3578 


\bibitem[Ziparo et al.(2012)]{2012MNRAS.420.2480Z} Ziparo, F., Braglia, 
F.~G., Pierini, D., et al.\ 2012, \mnras, 420, 2480 
\end{thebibliography}
\end{document}